\documentclass[12pt,preprint]{aastex}
\usepackage{amsmath}
\usepackage{amssymb}

\newcommand{\etal}{et~al.~}

\newcommand{\kms}{\ifmmode\,{\rm km}\,{\rm s}^{-1}\else km$\,$s$^{-1}$\fi} 
\newcommand{\be}{\begin{equation}}
\newcommand{\ee}{\end{equation}}
\newcommand{\bea}{\begin{eqnarray}}
\newcommand{\eea}{\end{eqnarray}}

\def \spose#1{\hbox to 0pt{#1\hss}}                                   
\def \ltsim{\mathrel{\spose{\lower 3pt\hbox{$\sim$}}                  
     \raise 2.0pt\hbox{$<$}}}                                                 
\def \gtsim{\mathrel{\spose{\lower 3pt\hbox{$\sim$}}                   
     \raise 2.0pt\hbox{$>$}}}

\def\se#1{\S\ref{sec:#1}}
 
\def\Fig#1{Figure~\ref{fig:#1}}

\def\ifm#1{\relax\ifmmode#1\else$\mathsurround=0pt #1$\fi}  
\def\kms{\ifmmode\,{\rm km}\,{\rm s}^{-1}\else km$\,$s$^{-1}$\fi}

\def\ltsima{$\; \buildrel < \over \sim \;$}                    
\def\lsim{\lower.5ex\hbox{\ltsima}}  
\def\gtsima{$\; \buildrel > \over \sim \;$}                    
\def\gsim{\lower.5ex\hbox{\gtsima}}

\def\C28{\rm C_{28}} 

\def\pmb#1{\setbox0=\hbox{#1}%
\kern-.025em\copy0\kern-\wd0 \kern.05em\copy0\kern-\wd0 
\kern-.025em\raise.0433em\box0} 
 
\def \ion#1#2{#1{\footnotesize{#2}}\relax}  
  
\def \hi {\ion{H}{I} }

\def \littlemm{\ifmmode{\scriptscriptstyle m } 
     \else{\hbox{$\scriptscriptstyle m $ }}\fi}  
\def \topemm{\raise .9ex \hbox{\littlemm}}  
\def \magpoint{\hbox to 2pt{}\rlap{\hskip -.5ex 
     \topemm}.\hbox to 2pt{}} 
 
\def \magarc {mag arcsec$^{-2}$}

\usepackage{lscape}
\usepackage{apjfonts}

\slugcomment{For submission to MNRAS. Not for distribution} 

\shorttitle{Formation and Evolution of Virgo Cluster Galaxies - I. Colours} 
\shortauthors{Roediger \etal 2011}

\begin{document}


\title{The Formation and Evolution of Virgo Cluster Galaxies -- \\I. Broadband Optical \& Infrared Colours}

\author{Joel C. Roediger and St\'ephane Courteau}    
\affil{Department of Physics, Engineering Physics \& Astronomy, Queen's 
University, Kingston, Ontario, Canada}

\author{Michael McDonald}  
\affil{Dept. of Astronomy, University of Maryland, College Park, MD}

\and

\author{Lauren A. MacArthur}
\affil{Herzberg Institute of Astrophysics, National Research Council of Canada, 
5071 West Saanich Road, Victoria, BC, Canada}

\email{jroediger,courteau@astro.queensu.ca,mcdonald@astro.umd.edu,
lauren.macarthur@nrc-cnrc.gc.ca}


\begin{abstract}
We use a combination of deep optical (gri) and near-infrared (H) photometry to 
study the radially-resolved colours of a broad sample of Virgo cluster 
galaxies. For most galaxy types, we find that the median g-H colour gradient is 
either flat (gas-poor giants and gas-rich dwarfs) or negative (i.e., colours 
become bluer with increasing radius; gas-poor dwarfs, spirals, and gas-poor 
peculiars). Also, later-type galaxies typically exhibit more negative gradients 
than early-types. Given the lack of a correlation between the central colours 
and axis ratios of Virgo spiral galaxies, we argue that dust likely plays a 
small role, if at all, in setting those colour gradients. We search for 
possible correlations between galaxy colour and photometric structure or 
environment and find that the Virgo galaxy colours become redder with 
increasing concentration, luminosity and surface brightness, while no 
dependence with cluster-centric radius or local galaxy density is detected 
(over a range of $\sim$2~Mpc and $\sim3-16$ Mpc${-2}$, respectively). However, 
the colours of gas-rich Virgo galaxies do correlate with neutral gas 
deficiency, such that these galaxies become redder with higher deficiencies. 
Comparisons with stellar population models suggest that these colour gradients 
arise principally from variations in stellar metallicity within these galaxies, 
while age variations only make a significant contribution to the colour 
gradients of Virgo irregulars. This paper serves as an introduction to a 
detailed stellar population analysis of Virgo cluster galaxies by the same 
authors (Roediger et al 2011b).
\end{abstract}

\keywords{galaxies: clusters: individual (Virgo) ---
          galaxies: dwarf ---
          galaxies: elliptical --- 
          galaxies: lenticular --- 
          galaxies: evolution ---
          galaxies: formation ---
          galaxies: irregular ---
          galaxies: peculiar ---
          galaxies: spiral ---
          galaxies: stellar content ---
          galaxies: structure}


\section{INTRODUCTION}\label{sec:Intro}

The broadband colours of galaxies reveal information about the ages and 
metallicities of their stellar populations (as well as their dust content). 
Galaxy colours thus provide a fundamental means for the investigation of both 
their star formation histories and chemical evolution, a fact which is 
reflected by the rich history of this field 
\citep[e.g.,][]{Va88,Pe90a,Go94,dJ96,BdJ00,Ma04,vZ04,Ta05} since the pioneering 
works of \cite{Se73} and \cite{TG76}.

The alternative to colour-based analyses of galaxies' stellar populations 
involves a spectroscopic approach, either through the measurement of absorption 
line strengths \citep[e.g.,][]{Fa73,Te81,Bu84,Wo92,KA99,Tr00,Gal05,Pe07} or the 
synthesis of entire spectra \citep[e.g.,][]{Ma09}. The benefits of the 
spectroscopic approach include a cleaner separation of the correlated age and 
metallicity effects \citep{Wo94} and, in the case of the Lick indices, a global 
insensitivity to dust reddening \citep{Ma05}. Spectroscopic stellar population 
studies demand some appreciable trade-offs compared to colour-based analyses 
however, such as smaller sample sizes and shallower depths. Colours then offer 
the most efficient means to simultaneously obtain deep radially-resolved and 
representative stellar population information for large numbers of galaxies.

Through comparison with stellar population models, galaxies' colour gradients 
may be decomposed into corresponding gradients of age and metallicity. The 
latter data provide a record of the build-up of galaxies' stellar contents and 
so are naturally of great value to galaxy formation and evolution models. In 
modelling galaxy colours, a wide and well-sampled wavelength baseline 
\citep[i.e., ultraviolet, UV, to infrared;][]{Ga02} is certainly desirable, not 
only to better constrain the fits, but also to overcome the age-metallicity 
degeneracy which plagues optical stellar population studies \citep{Wo94}. 
Although the availability of space-based imaging (i.e., GALEX) has provided a 
much desired wavelength expansion of colour-based studies 
\citep[e.g.,][]{Bo08}, the use of optical and near-infrared (NIR) data alone 
already goes a long way towards overcoming the above degeneracy \citep{Ca03}.

Radially-resolved studies of galaxy colours abound in the literature. For giant 
gas-poor galaxies (E/S0), the consensus seems to be that they possess negative 
colour gradients \citep{PV89,Pe90a,Go94,SE94,Mi99,Wu05,Ja06,LBdC09,GP10,LB10}, 
the cause of which is popularly attributed to an outward decrease in stellar 
metallicity \citep{Va88,Fr89,Ta00,TO00,TO03,LB04}. Positive UV and optical 
colour gradients found in both local and high-$z$ E/S0's 
\citep{Mi00,ME02,Me04,Fe09,Su10}, however, imply that centralized star 
formation has recently occurred, thereby arguing for at least a partial age 
gradient contribution. 

Discerning the nature of spiral galaxies' colour gradients is inherently 
complicated due to the presence of dust, gas, and multiple stellar populations 
within them \citep{KA90,GdA01,Ko05,Ta05,Ga09,Li09}. This complexity may be 
lifted though by distinguishing between early- and late-type spirals, wherein 
the former have flat or negative colour gradients 
\citep{Te94,dJ96,PB96,Mo01,Ga09} while the latter are described by positive 
colour gradients, but with a wide dispersion \citep{KA90,Ta05,TT06,Li09}. Given 
such a variety, it is not surprising that spiral galaxies' colour gradients
 have often been ascribed to both age and metallicity effects, with dust making 
a second-order or altogether negligible contribution 
\citep[; however, see \citealt{Ga09}]{dJ96,PB96,GdA01,Ma04}.

Lastly, the colour gradients amongst the remaining galaxy types are more 
uniform than those found in spirals. That is, both peculiar and (gas-rich and 
gas-poor) dwarf galaxies are typically described by positive colour gradients 
\citep{Va88,Je00,KA90,Ja94,Pa96,Tu96,Ba03,vZ04,Ta05}. Since the notion of such 
galaxies having a higher metallicity in their outskirts (where the potential is 
weaker) seems highly counter-intuitive, these colour gradients have commonly 
been attributed to underlying positive age gradients within them\footnote{The 
effects of dust and recent star formation may, however, also contribute to the 
positive colour gradients in gas-poor dwarfs.} \citep[e.g.,][]{Va88}. Negative 
metallicity gradients, on the other hand, may very well contribute to the flat 
and negative colour gradients that have been observed within a minority of 
dwarf galaxies \citep{PT96,Pa02,Ba03,vZ04}.

Aside from the typical colour gradient for each principal morphological type, 
the structural or environmental parameters which correlate with galaxies' 
colours and colour gradients are also of interest to understanding galaxy 
formation. The most prevalent of such correlations for all galaxy types are 
those against luminosity \citep{BP94,Ja00,TO03,MM07,Li09,To10a}, mass 
\citep{Va88,Ma04,We09b,LB10,Su10}, colour \citep{Tu96,Pi02,GdA01,Fe05,Fe09}, 
and size \citep{Be99,Pa02,TO03,MM07,Li09,To10a}, in the sense that more 
luminous, more massive, redder, and larger galaxies have reddest colours and 
flattest colour gradients. In addition, spiral galaxies have bluer colours and 
steeper colour gradients for lower surface brightnesses \citep{dJ96,Li09} and 
larger \hi contents \citep {We09b}. Less uniformity has been found amongst 
galaxy types with respect to the environmental dependence of their colour 
information, however. While the colour gradients of gas-poor giants flatten 
\citep{KI05,LB05} and the colours of gas-rich dwarfs redden \citep{GH86,Pa02} 
towards high densities, no corresponding trends have been found amongst spiral 
galaxies \citep{Ko05,MM07}.

Despite the seemingly coherent picture presented above based upon piecemeal 
information, a homogeneous analysis of galaxy colours and colour gradients 
across all galaxy types which incorporates NIR data would enable a valuable 
confirmation of that picture. Thus, we present here the resolved optical and 
optical-NIR colours of a sample of Virgo cluster galaxies that spans the full 
range in galaxy morphology. This paper is organized as follows. In \se{Data} we 
discuss the salient aspects and organisation of our photometric database. 
Broadband colour gradients, as well as morphological, structural, and 
environmental trends with galaxy colours are shown and discussed in the context 
of previous work in \se{R&D}. The interpretation of our results in terms of the 
stellar populations of Virgo cluster galaxies, however, is addressed in a 
companion paper (Roediger et al. 2011b; hereafter, Paper II). Conclusions are 
provided in \se{Concs}. Throughout this paper, we assume a distance of 16.5 Mpc 
to all Virgo cluster galaxies \citep{Me07} and refer to both S0 and S0/a galaxy 
types simply as ``S0''.


\section{DATA}\label{sec:Data}

A systematic study of the colours and stellar populations within galaxies of 
all types demands as complete a volume-limited sample as possible. Given that 
the Virgo cluster is the closest large concentration of galaxies whose members 
span the full range of galaxian parameter space, we use for our study the 
radially-resolved $griH$ photometric database derived from the imaging survey 
of Virgo galaxies of McDonald et~al. (2011, hereafter M11). Only those aspects 
of M11 which are relevant to our study are discussed below; further details can 
be found in M11.

The M11 sample includes 285 galaxies selected from the Virgo Cluster Catalog 
\citep[hereafter VCC]{Bi85}; it is nearly complete down to $B \sim$ 16 ($M_B 
\sim$ -15). For each galaxy, Sloan Digital Sky Survey 
\citep[hereafter SDSS]{Ad07,Ad08} $gri-$ and separate $H-$band imaging is 
available. The spatial distribution and completeness of this sample are shown 
in Figures 1 \& 2 of M11. The M11 sample extends out to the 6$^{\circ}$ radius 
of the Virgo cluster \citep{Mc99}, such that the sample spans a wide range in 
environmental conditions (i.e., galaxy densities $\Sigma \sim$ 3$-$16 
Mpc$^{-2}$ and gas deficiencies $Def_{\hi} \sim$ -0.4$-$2.3 dex; see 
\se{Enviro}. As seen in M11, a natural bias for gas-poor galaxies exists in the 
sample given the morphology-density relation \citep{Dr80}. Nevertheless, the 
M11 sample contains 86 gas-rich galaxies (56 giants and 30 dwarfs; see Table 
\ref{tbl:Cgrd}), thereby allowing for meaningful comparisons of the colours, 
and thus stellar populations, between major Virgo galaxy types.

\subsection{Photometry}\label{sec:Phot}
Radially-resolved optical ($ugriz$) photometry for the M11 galaxies was 
extracted from SDSS imaging. However, the $u$- and $z$-bands were excluded from 
our analysis in light of their low signal-to-noise (hereafter $S/N$) and the 
significant number of light profiles having obvious sky subtraction errors at 
those wavelengths (M11). The M11 NIR photometry came from a mixture of imaging 
sources, including the GOLDMine \citep[78 galaxies]{Ga03} and Two Micron 
All-Sky Survey \citep[20 galaxies]{Sk06} databases, while $H$-band imaging for 
the remaining (187) galaxies was obtained by M11 from April 2005 to May 2008 on 
Mauna Kea telescopes [ULBCAM (UH 2.2 m); WIRCAM (CFHT); WFCAM (UKIRT)]. The 
spatial extent of the NIR light profiles is, on average, $\sim$60\% that of the 
corresponding $i$-band profiles for $\sim$200 galaxies (see \se{SBPP} for our 
definition of profile extent), while the optical and NIR profiles reach median 
surface brightness levels of $r \sim$ 26.5 \magarc\ and $H \sim$ 23 \magarc, 
respectively. The high quality of the M11 light profiles enables a reliable 
investigation of Virgo galaxies' colours out to typically large galactocentric 
radii.

\subsection{Surface brightness profiles and structural diagnostics}
\label{sec:SBP&SD}
The $griH$ major-axis surface brightness (hereafter SB) profiles of Virgo 
galaxies were extracted by M11 with the astronomical image-processing software 
XVISTA\footnote{\emph{http://ganymede.nmsu.edu/holtz/xvista/}}. To ensure that 
colour gradients are computed homogeneously, the SDSS SB profiles were degraded 
to the (poorer) resolution of the NIR profiles. 

We have also drawn upon the collection of non-parametric, $H$-band structural 
parameters presented in \citet{Mc09} to search for possible correlations 
between galaxy colour and structure. The structural parameters of interest 
include effective surface brightnesses ($\mu_{e}$), effective radii ($r_{e}$),
concentrations ($C_{28}$) and apparent magnitudes ($m_H$). These non-parametric
diagnostics all involve the extrapolation and integration of galaxies' NIR light
profiles to infinity. We use the NIR values of the above structural diagnostics
for their lower sensitivity to dust attenuation and greater sensitivity to the
galaxy stellar masses. 

\subsection{Environmental diagnostics}\label{sec:Enviro}
To probe for an environmental dependence in the Virgo galaxy colours, we 
characterize the impact of environment with the neutral hydrogen deficiency 
parameter, $Def_{\hi}$. For a cluster galaxy with a given gas mass $M_{\hi}$, 
$Def_{\hi} \equiv \left\langle \log M_{\hi}(D_{opt,T}) \right\rangle$ - $\log 
M_{\hi}$, where $\left\langle \text{log} M_{\hi}(D_{opt,T}) \right\rangle$ is the 
typical \hi mass for field galaxies of the same Hubble type, $T$, and optical 
diameter, $D_{opt}$, such that positive $Def_{\hi}$ values correspond to a dearth 
of neutral gas. The advantage of $Def_{\hi}$ as a measure of galaxy environment 
is its independence of projection effects that plague other traditional (2D) 
diagnostics \citep[hereafter S02]{So02}. For completeness, we still appeal to 
the (projected) cluster-centric distances, $D_{M87}$, and local surface 
densities of the M11 galaxies, $\Sigma$, to assist our study of environmental
effects on their colours. Cluster-centric distances were computed relative
to M87 while assuming a common distance to all VCC galaxies of 16.5 Mpc 
\citep{Me07}. We follow \citet{Dr80} in computing $\Sigma$ as the number 
density comprised by the ten nearest VCC neighbours to a given galaxy's 
position.

\begin{figure*}
 \begin{center}
  \begin{tabular}{c c}
   \includegraphics[width=0.45\textwidth]{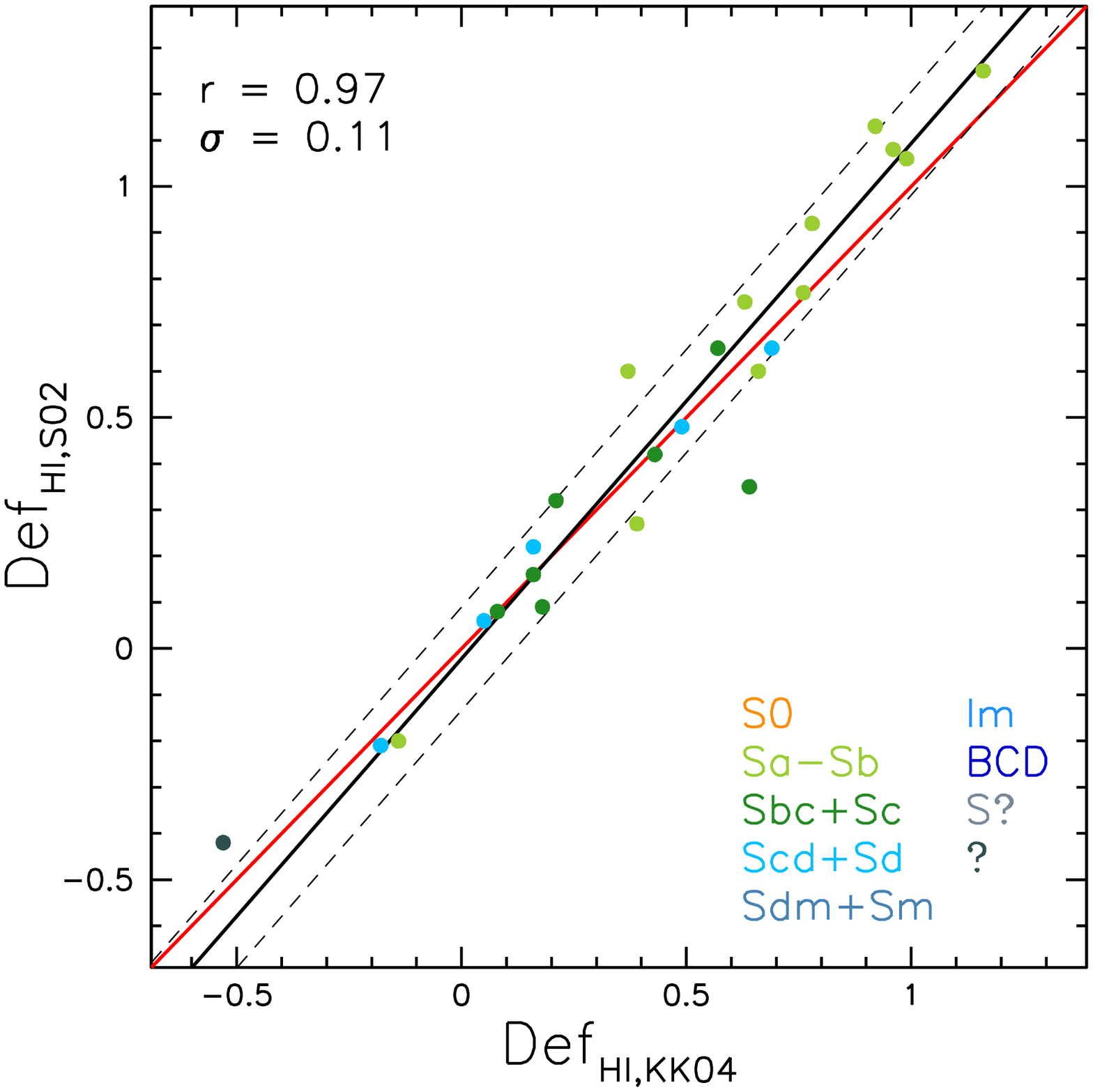} &
   \includegraphics[width=0.45\textwidth]{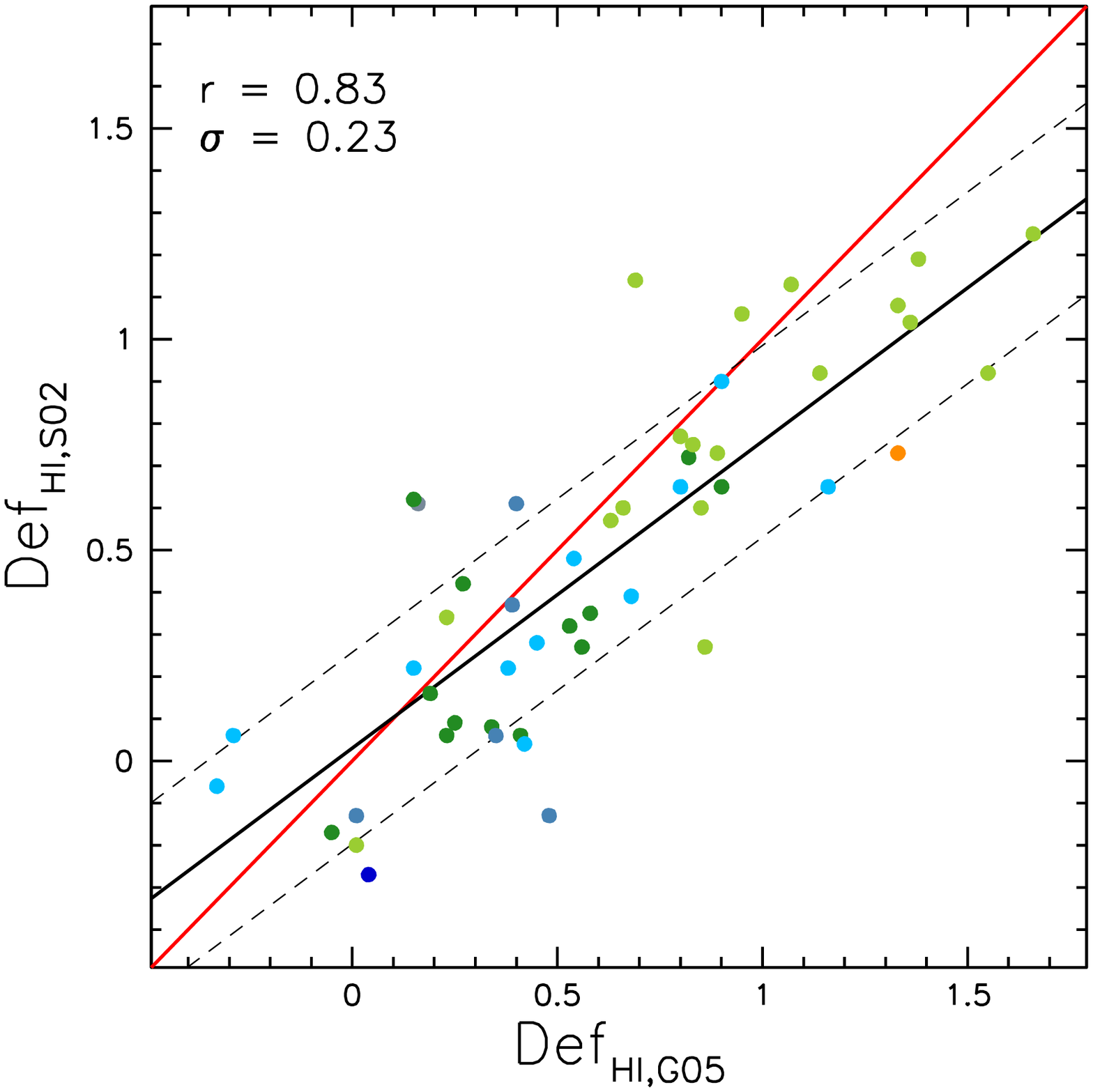}
  \end{tabular}
  \caption{(\emph{left}) Comparison of $Def_{\hi}$ measurements for Virgo 
galaxies in common between S02 and KK04. The red line has slope unity, while 
the solid and dashed black lines represent a linear fit to the data and the 
1$\sigma$ deviation about the fit, respectively. The Pearson (linear) 
correlation coefficient $r$ and $\sigma$ of the fit are both provided in the 
top-left corner. (\emph{right}) Same as left, but comparing S02 and G05.}
  \label{fig:DefHI-Comp}
 \end{center}
\end{figure*}

We obtained $Def_{\hi}$ measurements for M11 galaxies from S02, 
\citet[hereafter KK04]{KK04}, and \citet[hereafter G05]{Gav05}. The latter 
compilation is the most complete, providing \hi mass measurements for 91 of the 
115 gas-rich galaxies in the M11 sample, as well as an additional 10 E/S0 
systems. The other \hi mass compilations have $Def_{\hi}$ determinations for 28 
(KK04) and 53 (S02) galaxies, only four of which are not included in G05. We 
compare $Def_{\hi}$ measurements in \Fig{DefHI-Comp} for galaxies in common for 
KK04 and G02 with S02, with the red line in each plot denoting equality. While 
there is good consistency between the KK04 and S02 databases ($r$ = 0.97; 
$\sigma$ = 0.11 dex), the measurements of G05 are systematically higher on 
average than the others, regardless of morphology. Presuming that the G05 
measurements are at least internally consistent, we rely strictly on that 
database for $Def_{\hi}$ values.

\subsection{Surface brightness profile preparations}\label{sec:SBPP}
Prior to measuring Virgo galaxy colours and stellar populations, the M11 SB 
profiles require special attention, which we now describe. The computation of 
accurate galaxy colour profiles relies on accounting for pixels that are 
spoiled by either foreground stars, scattered light, or low $S/N$. To correct 
for this, we first truncated the M11 SB profiles by searching outward in each 
of them for a sequence of adjacent isophotes with SB errors in excess of a 
given threshold $\sigma_{\mu,th}$. The isophotes beyond that point were flagged 
as unreliable and their location marked as the truncation radius.

In the truncation process, we used error thresholds of 0.25 and 0.20 \magarc\
for the M11 optical and NIR SB profiles, respectively. These values correspond 
to locations in the SB profiles where the point-to-point scatter becomes 
comparable to the statistical errors per isophote. These thresholds are also 
roughly equal to 0.1-0.2\% of the sky level (or less) in each profile.

We have also applied an adaptive radial binning scheme to the M11 SB profiles 
to increase the $S/N$ per radial bin. While previous studies of galaxy colour 
gradients \citep[e.g.,][]{Ma04} used an incremental radial binning scheme, ours 
requires that a minimum signal-to-noise $S/N_{\text{min}}$ be met in all four 
bands in a given bin. Since an incremental scheme makes no such stipulation and 
typically has larger radial bins, the colour profiles of individual galaxies
will then tend to extend further in this scheme than those determined from 
our adaptive scheme. However, these increased depths are overshadowed by the 
higher colour dispersions per radial bin and the inclusion of low $S/N$ data 
points in galaxies' outskirts, thereby decreasing the quality of the colour 
profiles overall.

Accounting only for the Poisson noise of the source and sky, we calculate the 
$S/N$ per isophote according to,
\begin{align}
 \dfrac{S}{N} & = \dfrac{S_{\text{gal}}}{\sqrt{N_{\text{gal}}^2 + N_{\text{sky}}^2}} \notag \\
              & = \dfrac{\mu_{\text{gal}} \times C_{\text{ell}}}{\sqrt{\mu_{\text{gal}} \times C_{\text{ell}} + \mu_{\text{sky}} \times C_{\text{ell}}}} \notag \\
              & = \dfrac{\mu_{\text{gal}}}{\sqrt{\mu_{\text{gal}} + \mu_{\text{sky}}}} \times \sqrt{C_{\text{ell}}},
 \label{eqn:S2N}
\end{align}
where $\mu_{\text{gal}}$ and $\mu_{\text{sky}}$ are the isophote's SB and sky 
background, respectively, and $C_{\text{ell}}$ is its circumference.

The radial binning scheme for each galaxy is based on the SB profile with the 
lowest overall $S/N$; the latter usually corresponds to the $g$-band, followed 
by the $H$-band. We enforce a minimum bin size for each galaxy, which is set by 
the maximum seeing disk amongst all bands and most often corresponds to the 
$H$-band value. The SB of a bin is determined by a weighted average of the 
isophotal SBs that it contains, while the bin location is set to its midpoint. 
Through extensive testing, we have found that using the $\sigma_{\mu}-\mu$ 
median trends determined from the original photometry provided the most 
reliable way of estimating the SB error per bin. Lastly, we have carried out 
our analysis for the cases of $S/N_{\text{min}}$ = 10, 20, 30, 40 and 50 per bin. 

\subsection{Sky errors}\label{sec:DSky}
The determination of accurate colour gradients for a galaxy demands reliable 
subtraction of the sky's contribution to the flux measured within each filter. 
An under- or over-estimated sky in a given band will artificially increase or 
decrease the corresponding SB measured in a galaxy's outskirts, respectively, 
thereby skewing its colour profiles. Sky errors dominate all other error 
sources in the study of galaxy colours \citep{BdJ00}. While M11 took great care 
in assessing sky level corrections, our profile truncation algorithm 
(\se{SBPP}) is already effective before the sky error envelopes become 
prohibitively large (i.e., $\sigma_{\text{sky}} \sim$ 0.2\% of the sky level). 
We revisit the issue of sky errors in \se{CP&G}.


\section{RESULTS \& DISCUSSION}\label{sec:R&D}

We now consider the broadband colour profiles of Virgo galaxies, as well as 
seek out possible morphological, structural and environmental trends in their 
central colours. These data offer a view of stellar populations both within and 
amongst galaxies, free of any assumptions about their underlying star formation 
histories and chemical evolution.

\subsection{Colour profiles and gradients}\label{sec:CP&G}
\begin{figure*}
 \begin{center}
  \includegraphics[width=0.9\textwidth]{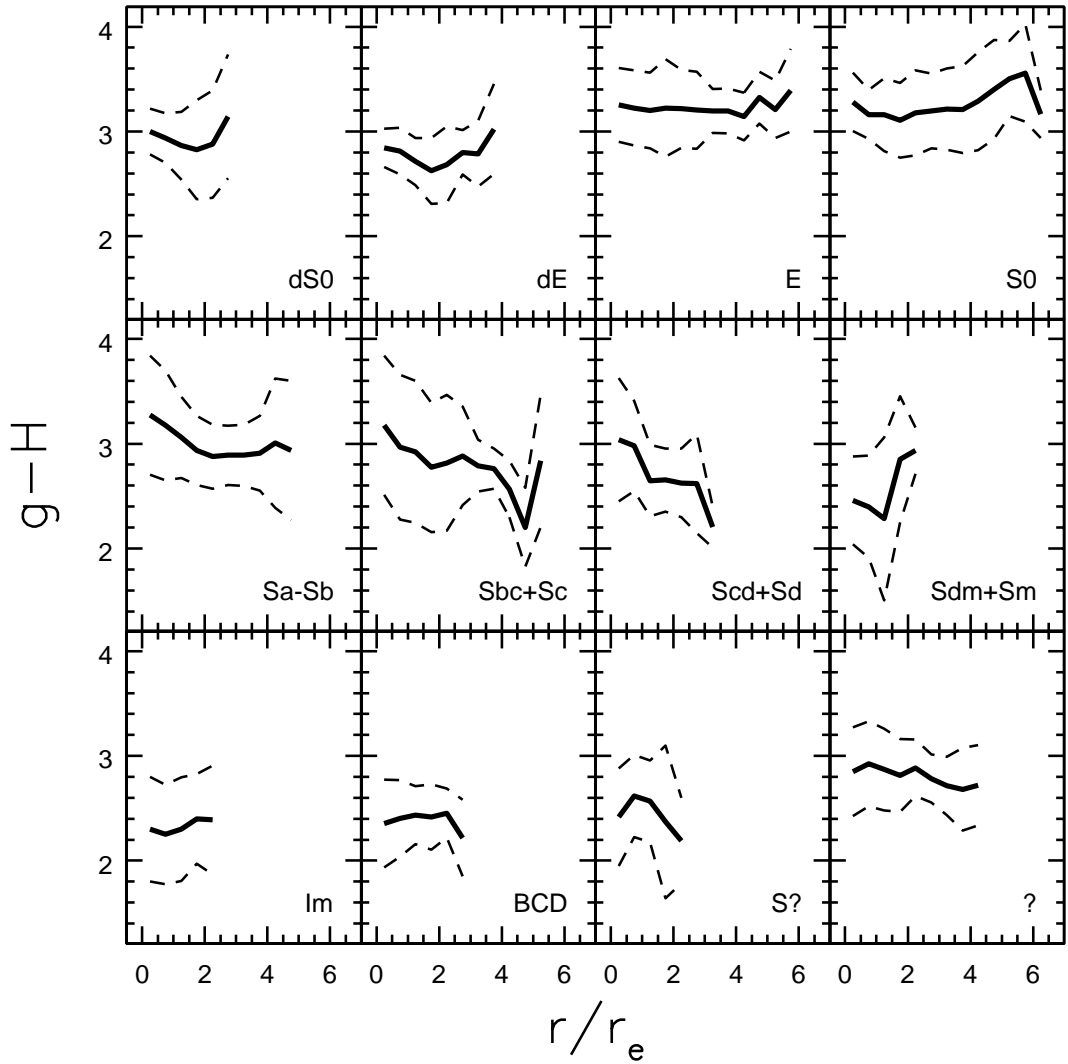}
  \caption{$g$-$H$ profiles as a function of scaled radius, $r/r_e$, for Virgo 
galaxies binned by morphological type. The median profiles and their 
1$\sigma$-dispersion are indicated by the solid and dashed lines, respectively.}
  \label{fig:gHprf}
 \end{center}
\end{figure*}

The wide frequency baseline of the $g$-$H$ colour provides us with our best 
constraint on the broad spectral shapes of Virgo galaxies, and thus on their 
stellar populations. The median $g$-$H$ profiles and rms dispersions for Virgo 
galaxies of different morphologies are shown in \Fig{gHprf} by the solid and 
dashed lines, respectively. For this exercise, the $g$-$H$ colour profile for
each galaxy was created by subtracting its binned $H$-band light profile from 
its binned $g$-band light profile. We also scaled each colour profile according 
to the NIR effective radius of the galaxy, thereby removing the size variation 
that exists amongst galaxies of various Hubble types. Given our imposed $S/N$ 
constraints, we find that the median colour profiles for the many Virgo galaxy 
types span a large range, extending out to $\sim$6 $r_e$ for the gas-poor 
giants, down to $\gtrsim$2 $r_e$ for the gas-rich dwarfs and irregulars. 

As a complement to \Fig{gHprf}, we list in Table \ref{tbl:Cgrd} the relevant 
colour information for each morphological group (column 1). Columns 2-4 contain 
the median $g$-$H$ colour gradients, where the gradient for each galaxy was 
measured from a linear fit to the $g$-$H$ profile. The gradients were computed 
for three choices of radial coordinate, so that the median gradients could be 
expressed either in terms of scaled ($r_e$; column 2) or physical radii (the 
latter were computed on both a linear and logarithmic scale; columns 3-4). 
Column 5 shows the median $g$-$H$ central colours, while column 6 provides the 
number of galaxies in each morphological group. Both the central colours and 
colour gradients for each of our galaxies, along with their surface brightness 
profiles and bulge-disk decompositions, are available at the Centre de 
Donn\'ees astronomiques de Strasbourg and our own 
website\footnote{\emph{www.astro.queensu.ca/virgo}}.

\begin{figure*}
 \begin{center}
  \includegraphics[width=0.9\textwidth]{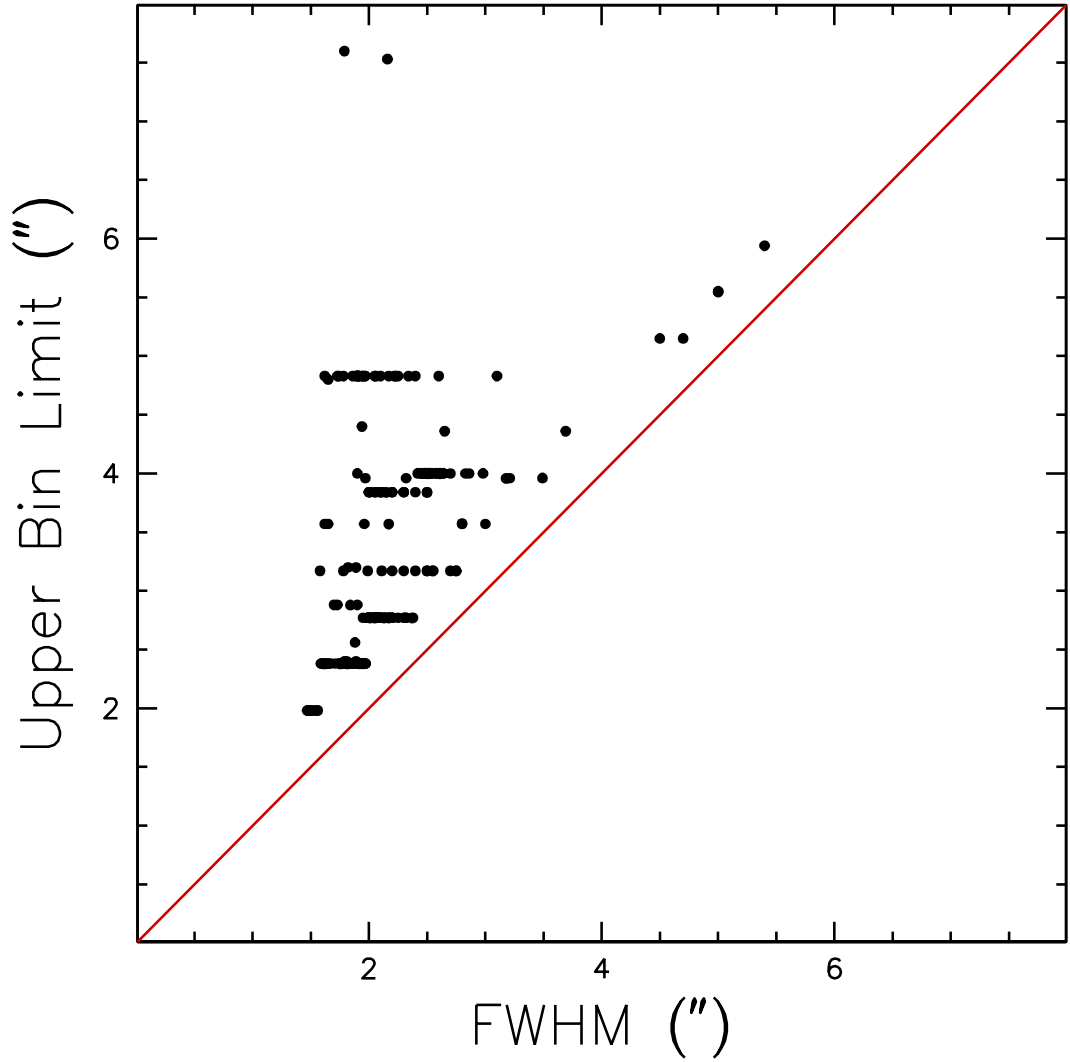}
  \caption{Size of the first radial bin versus the FWHM of the maximum seeing 
disk for Virgo galaxies. The 1:1 relation is marked by the red line.}
  \label{fig:BinSize}
 \end{center}
\end{figure*}

In fitting galaxies' colour profiles, we neglected the first radial bin in 
order to avoid adverse PSF blurring effects from skewing our computed 
gradients. We show in \Fig{BinSize} the size of the first radial bin versus the 
FWHM of the maximum seeing disk for all Virgo galaxies. The proximity of many 
galaxies to the line of equality (red) indicates that seeing effects within the 
first bin of the colour profiles for those systems are likely non-negligible. 

\begin{figure*}
 \begin{center}
  \includegraphics[width=0.9\textwidth]{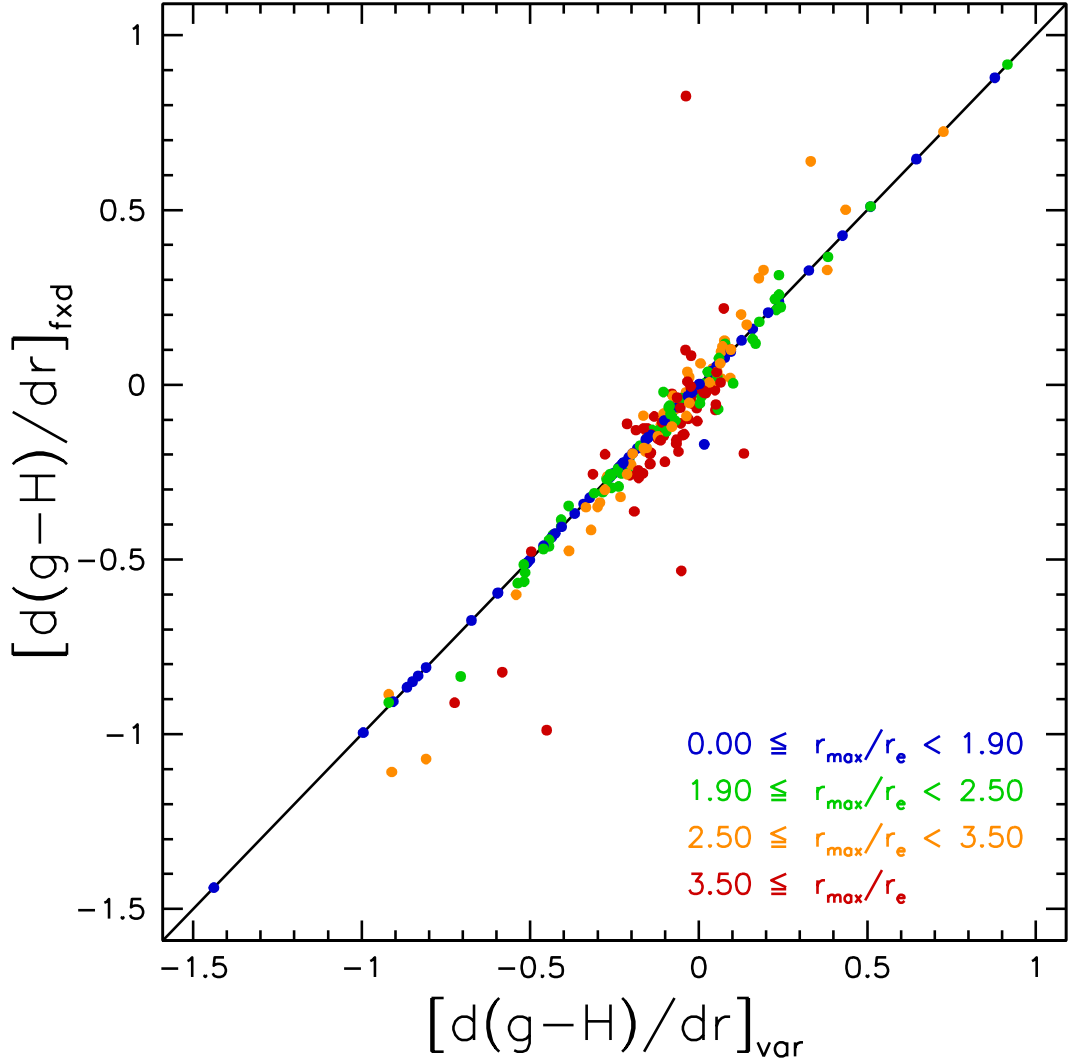}
  \caption{Comparison of the $g$-$H$ colour gradients for Virgo galaxies 
determined using either a fixed or variable radial range in the linear fits to 
the colour profiles of individual galaxies. The data point colours reflect
the maximum extent of each galaxy colour profile. For the fixed radial
range fits, only data points within $r <$~2$r_e$ were included. The 
black line marks the 1:1 relation.}
  \label{fig:FxdvsVar}
 \end{center}
\end{figure*}

\begin{figure*}
 \begin{center}
  \includegraphics[width=0.9\textwidth]{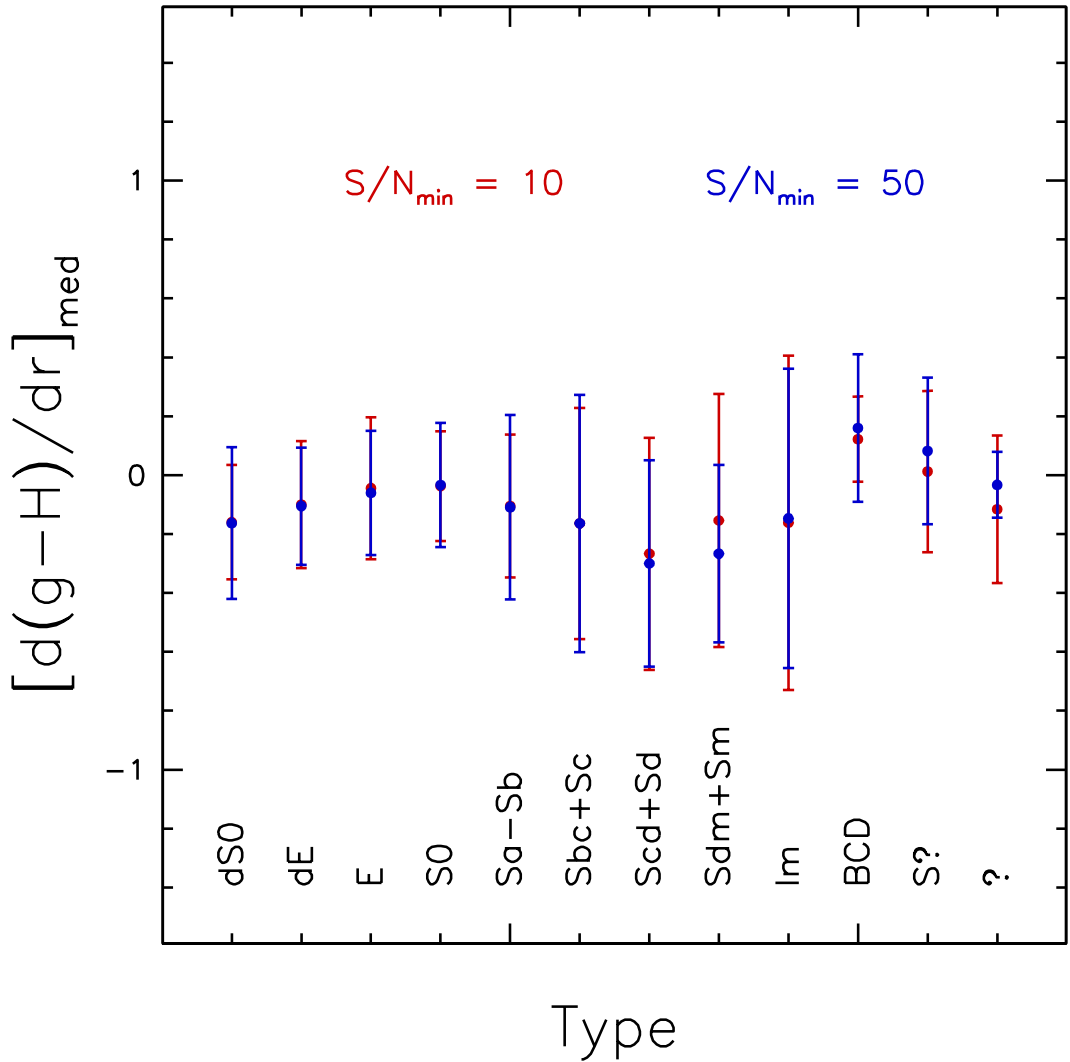}
  \caption{Comparison of the median $g$-$H$ colour gradient estimates (as 
determined in $r_e$ space) for Virgo galaxies, binned by morphological type, 
using a minimum $S/N$ of either 10 or 50. The size of the error bars reflects 
the 1$\sigma$ (rms) dispersion within each morphological bin.}
  \label{fig:SN10vsSN50}
 \end{center}
\end{figure*}

Other effects which could potentially skew the median colour gradient estimates 
provided in Table \ref{tbl:Cgrd} include the use of a variable radial range in 
individual colour profile fits (given our choice of binning scheme) and of 
different $S/N_{min}$ values in adaptively binning the M11 SB profiles. As shown 
in Figures \ref{fig:FxdvsVar} and \ref{fig:SN10vsSN50} though, neither of these 
two effects presents a significant risk towards the robustness of our colour 
gradient determinations.

\begin{figure*}
 \begin{center}
  \includegraphics[width=0.9\textwidth]{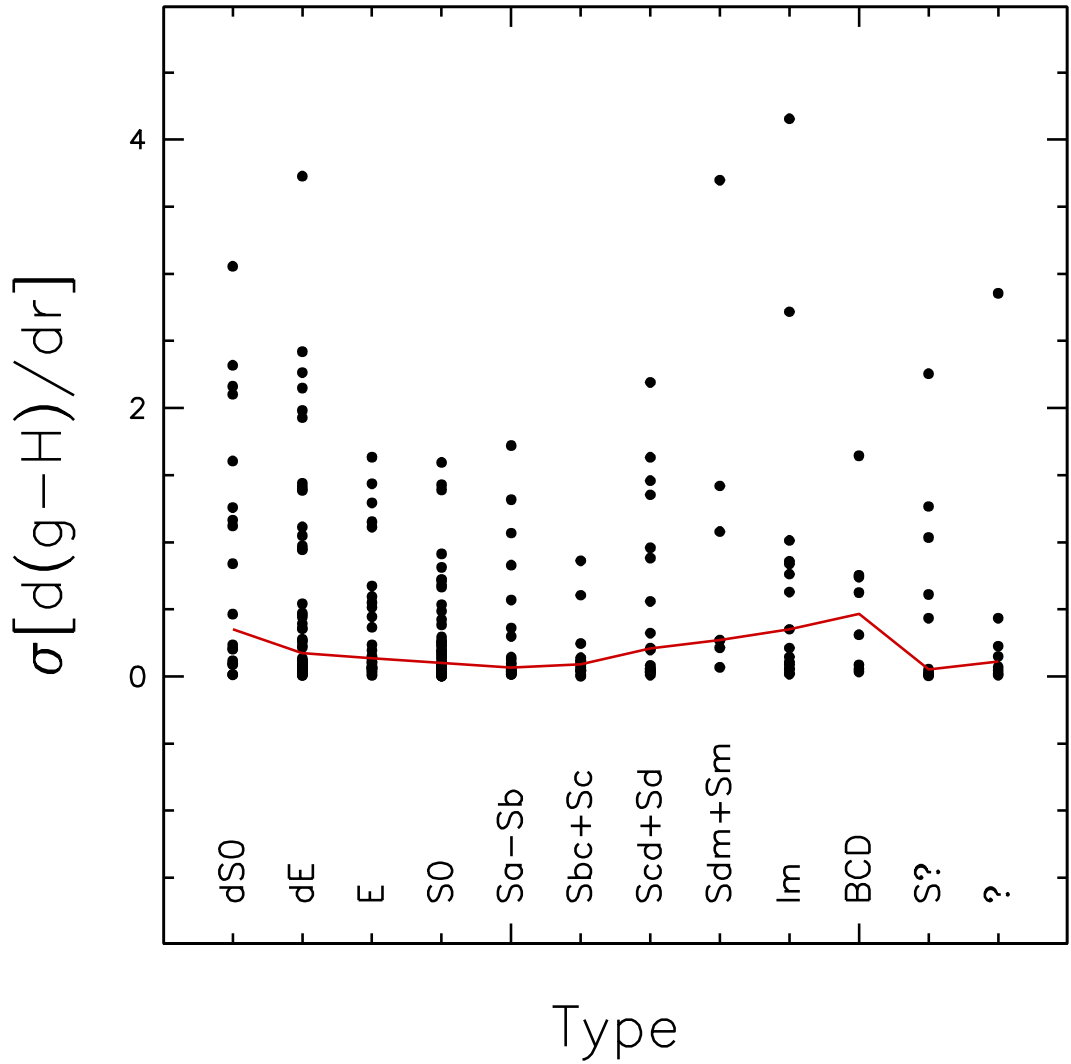}
  \caption{The (potential) contribution of sky subtraction errors to the 
estimated colour gradients (in $r_e$ space) of individual Virgo galaxies, 
binned by morphological type. The red line connects the median error within 
each morphological bin.}
  \label{fig:dSky}
 \end{center}
\end{figure*}

Examination of Table \ref{tbl:Cgrd} reveals a large dispersion in the colour 
gradient distribution of each Virgo galaxy type. These dispersions may be 
either intrinsic to Virgo cluster galaxies, or largely the result of 
measurement errors in their SB profiles. As mentioned in \se{DSky}, imperfect 
sky subtraction is the largest source of error in the computation of galaxy 
colour gradients. We can estimate the impact of sky errors on those gradients 
through the $\pm$1$\sigma_{sky}$ error envelopes for each M11 SB profile, where 
$\sigma_{sky}$ is the standard deviation in the sky levels measured in boxes 
along the periphery of a given Virgo galaxy image. We assess the colour 
gradient error by calculating the extrema of this quantity based on sky error 
envelopes. The distribution of errors in the $g$-$H$ colour gradients of Virgo 
galaxies due to imperfect sky subtractions is shown in \Fig{dSky}. Comparing 
\Fig{dSky} with Table \ref{tbl:Cgrd}, it can be seen that median sky errors 
contribute no more than $\sim$16\% of the dispersion found in the colour 
gradient distribution of each galaxy type; those dispersions are thus largely 
intrinsic to each galaxy population.

It is reassuring that the distribution of the median colour gradients listed in 
Table \ref{tbl:Cgrd} accurately reflects our impressions from \Fig{gHprf}, 
which we now describe\footnote{The apparent contradictory behaviour exhibited 
in the outskirts of some median colour profiles (e.g., dE) is resolved by 
considering that the colour data in the outskirts have lower $S/N$ than that 
which comprises the interiors}. Collectively, the median $g$-$H$ profiles of 
Virgo galaxies exhibit intriguing features. First, most galaxy types (dE/dS0, 
Sa$-$Sd, Im, ?) are described by a negative colour gradient, albeit of 
differing strengths. The gas-poor E/S0's, Sdm+Sm spirals, star-forming BCDs, 
and morphologically-peculiar S? galaxy types, on the other hand, have either 
quasi-flat (E/S0, Sdm+Sm, BCD) or positive (S?) colour gradients. This 
variation in the $g$-$H$ gradients of Virgo galaxies seems to correlate with 
morphology, such that the gradient is more negative for later Hubble types, as 
seen elsewhere \citep[e.g.,][]{Ja00,LB02,GP10}. Although the large dispersions 
reported in Table \ref{tbl:Cgrd} might draw suspicion on such a claim, 
comparisions of the gradient distributions for neighbouring morphologies via 
the Kolmogorov-Smirnov (KS) test confirms that the probability that any two 
distributions come from the same parent does not exceed $\sim$40\%.

On its own, our detected morphological trend in the median colour gradients of 
Virgo galaxies might seem surprising as previous studies of galaxy colours have 
failed to reveal the existence of such gradient (i.e., positive/negative) for 
most galaxy types \citep[e.g.,][]{KA90,GdA01,Gr05,HE06,Li09}. The dispersions 
in Table~\ref{tbl:Cgrd} show, however, that the colour gradient distribution 
for each Virgo galaxy type encompasses both positive and negative values, 
confirming that the variety in colour gradient sign found previously is 
reproduced here. Late-type spirals, on the other hand, have been shown by 
\citep{Ta05} as having a robust (positive) colour gradient sign, which, 
interestingly enough, is contrary to the typically negative gradient that we 
find for Virgo Scd$-$Sm spirals. This discrepancy may be attributed to the fact 
that Taylor et al. used UV-optical colours, which are quite sensitive to the 
ongoing star formation that is likely occurring within these gas-rich galaxies.

Also interesting from Table \ref{tbl:Cgrd} is the variation of the colour 
gradient dispersion amongst major galaxy types. For instance, the (rms) 
dispersion (in $r_e$ space) for the Virgo gas-poor giants and spirals is 
$\sim$0.21 and $\sim$0.36, respectively. The irregulars have intermediate 
values. This result agrees with the significant scatter in both the colours and 
colour gradients of spiral galaxies that has been reported by many 
\citep{PB96,Ta05,MM07,Ga09,We09b}. The large colour ranges spanned by both 
Virgo spirals and irregulars may reflect a large variance in their stellar 
population properties due to their ongoing star formation. Dust attenuation may 
also be affecting these galaxies' colours, especially since the M11 sample 
contains a few unrealistically red ($g$-$H \sim$ 4-5) Sa$-$Sd galaxies (see 
\Fig{CC-Sp}). Whether the role of dust in setting galaxy colours is 
statistically important or not will be revisited in \se{CT-Dust}.

In addition to the collective properties and trends noted above, inspection of 
the median colour profiles amongst major Virgo galaxy types reveals other 
noteworthy results. For instance, within their respective errors, the profiles 
for the two gas-poor giant types (E/S0) are remarkably similar; both exhibit a 
relatively constant colour ($g$-$H \sim$ 3.2) out to $\sim$4 $r_e$, beyond 
which they become redder with radius. This suggests that Virgo E/S0's possess 
similar stellar populations and likely share a similar (extended) formation 
history; this idea is explored further in Paper II. On the other hand, 
\Fig{gHprf} and Table~\ref{tbl:Cgrd} remind us that Virgo gas-poor dwarfs are 
systematically bluer and have more negative colour gradients than the gas-poor 
giants. Indeed, KS tests of the colour gradient distributions for Virgo
gas-poor dwarfs and giants only have a $\sim$7\% probability of having
a common parent distribution. The colour gradients that we have found for
Virgo gas-poor dwarfs are consistent with these galaxies being less enriched
overall than Virgo gas-poor giants (\citealt{Za94}; \citealt{Tr04}; Paper II).

\subsubsection{Comparison with literature}\label{sec:CGrd-Comp}
It is instructive to compare our optical-NIR colour gradient estimates with 
literature values for other galaxies. \cite{LB10} measured 
d($g$-$H$)/dlog(r/r$_e$) colour gradients in 4546 giant gas-poor galaxies that 
overlap with both the SDSS and 2MASS to find a strongly negative mean value and 
large dispersion for them: $-0.286\pm0.293$ mag/dex. The combined $g$-$H$ 
colour gradient distribution for our Virgo E and S0 galaxies, equal to 
$-0.245\pm0.426$ mag/dex, agrees well with La~Barbera's result. Moreover, the 
analysis of the g-K colour gradient distribution for a smaller sample of nearby 
giant gas-poor galaxies by \cite{Wu05} has also yielded a mean value similar to 
ours (-0.29 mag/dex).

With respect to gas-rich dwarf galaxies, \cite{HE06} analysed the $V$-$J$ 
colour gradients in both Magellanic irregulars and blue compact dwarfs and 
found that the mean colour gradient for these two galaxy types is -0.07 and 
+0.90 mag/kpc, respectively. While the mean $g$-$H$ colour gradient that we 
have estimated for Virgo Im galaxy types (-0.11 mag/kpc) agrees well with the 
former, that for Virgo BCDs (+0.11 mag/kpc) is much flatter. This discrepancy
is reduced when we consider the fact that the standard deviations about both
mean values is large (1.129 and 0.318 mag/kpc, respectively) and that both
distributions suffer from small sample sizes (N = 5 and 8, respectively).

Lastly, \cite{MM07} studied the $FUV$-$K$ colour gradients for spiral galaxies 
of all types. Although a quantitative comparison of FUV-$K$ and $g$-$H$ colours 
and colour gradients seems ill-conceived for any galaxy type, it is interesting 
to note that the median FUV-K colour gradient in these authors' sample 
decreases, albeit weakly, from early- to late-type spirals (-0.015 mag 
kpc$^{-1}$ for Sa$-$Sb's to -0.040 mag kpc$^{-1}$ for Scd+Sd's); their Sdm+Sm 
spirals do not follow this trend, however, having a nearly flat median gradient 
of -0.005 mag kpc$^{-1}$. This morphological trend in the Mu\~noz-Mateos et~al. 
data is similar to the one which we have found in terms of the $g$-$H$ colour 
gradients of Virgo cluster spirals. 

Based on these comparisons, encompassing a wide range of galaxy types, Virgo
cluster galaxies exhibit radial colour distributions that are apparently 
similar to those of field and cluster galaxies found in other nearby galaxy 
catalogs.

\subsection{Colour trends}\label{sec:CT}
\subsubsection{Morphology}\label{sec:CT-Morph}
\begin{figure*}
 \begin{center}
  \begin{tabular}{c c}
   \includegraphics[width=0.45\textwidth]{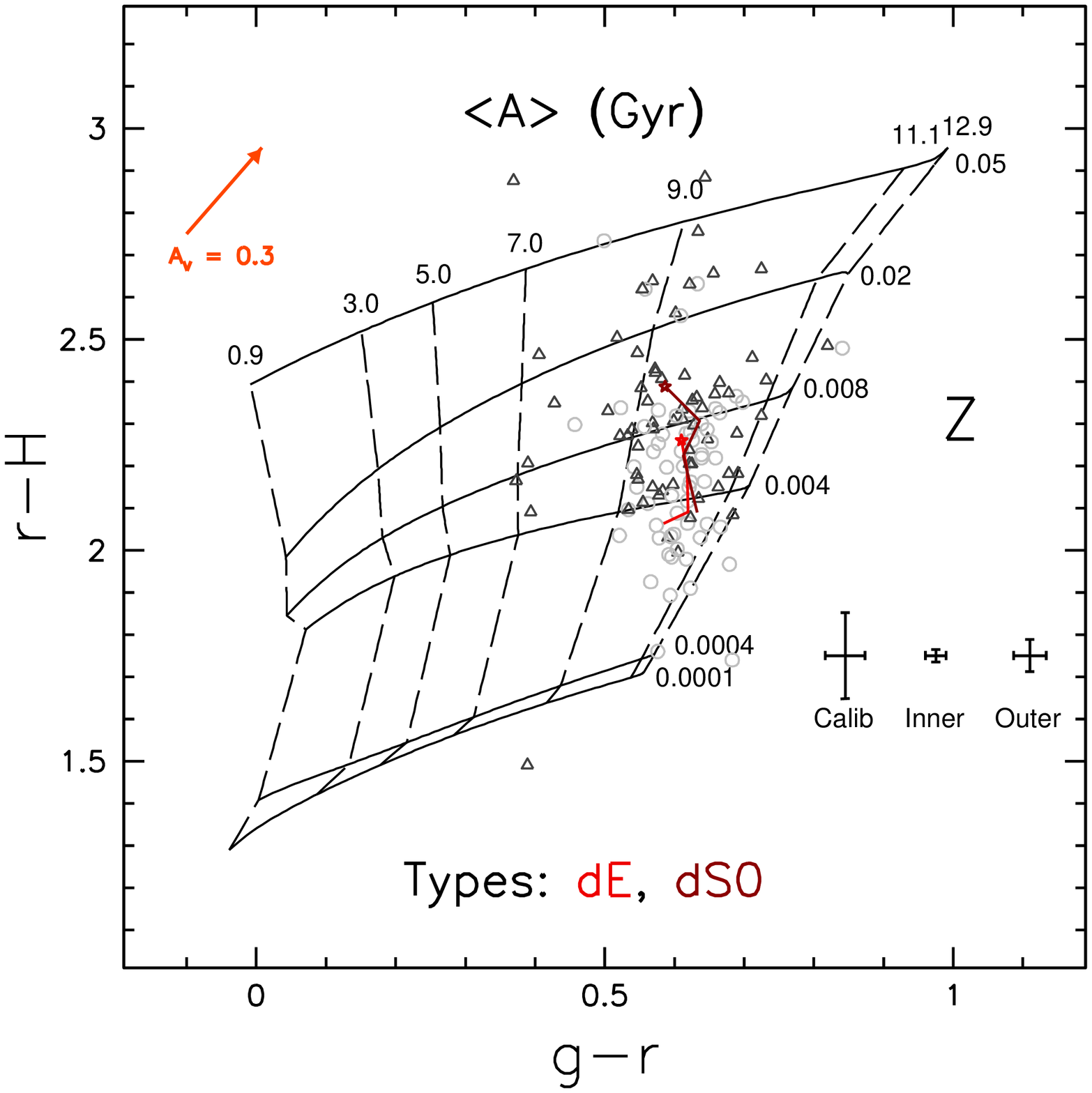} &
   \includegraphics[width=0.45\textwidth]{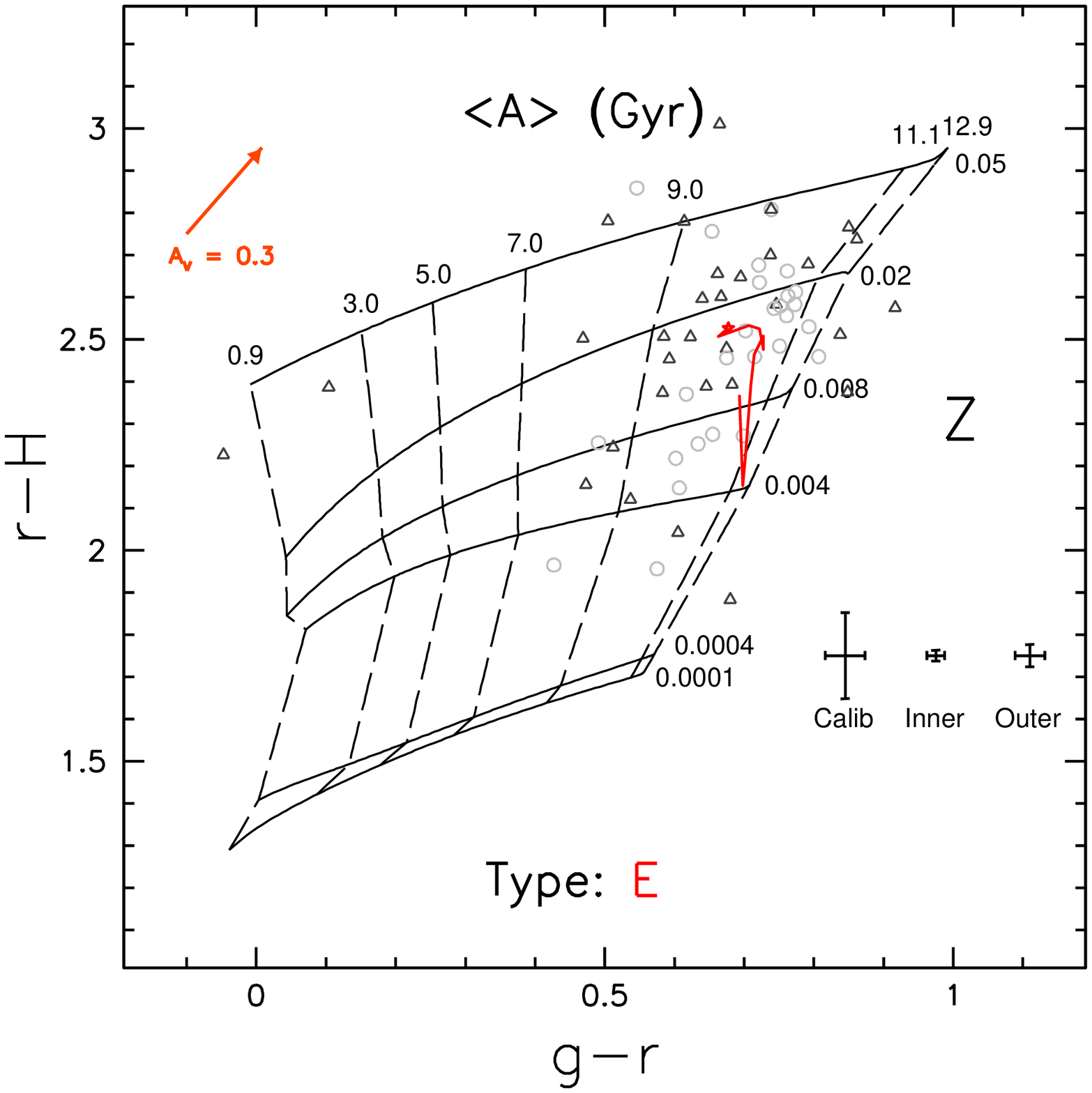}
  \end{tabular}
  \caption{(\textit{left}) $r$-$H$ versus $g$-$r$ colour-colour diagrams for 
Virgo dE and dS0 systems. Stars and filled circles represent the radial bins 
which enclose the galaxy centers and effective radii, respectively, 
while the coloured lines show the median colour profiles for both galaxy types. 
The grid represents a stellar population model for an exponential star 
formation history with no chemical evolution. The dashed and solid black lines 
correspond to lines of constant mean age and metallicity, respectively, with 
representative values indicated throughout the grid. The metallicities 
are expressed in percentages, according to mass contribution (i.e., $Z_{\odot}$ 
= 0.02). The error bars shown at lower-right indicate the typical uncertainties 
in the measured colours due to both calibration and sky subtraction errors for 
the entire sample (the latter is shown for both galaxies' central and outermost 
radial bins). The orange-red arrow in the top-left corner shows the reddening 
vector of a foreground dust screen model for the indicated amount of 
extinction. (\textit{right}) Same as left but for Virgo E galaxies.}
  \label{fig:CC-ETG}
 \end{center}
\end{figure*}

\begin{figure*}
 \begin{center}
  \begin{tabular}{c c}
   \includegraphics[width=0.45\textwidth]{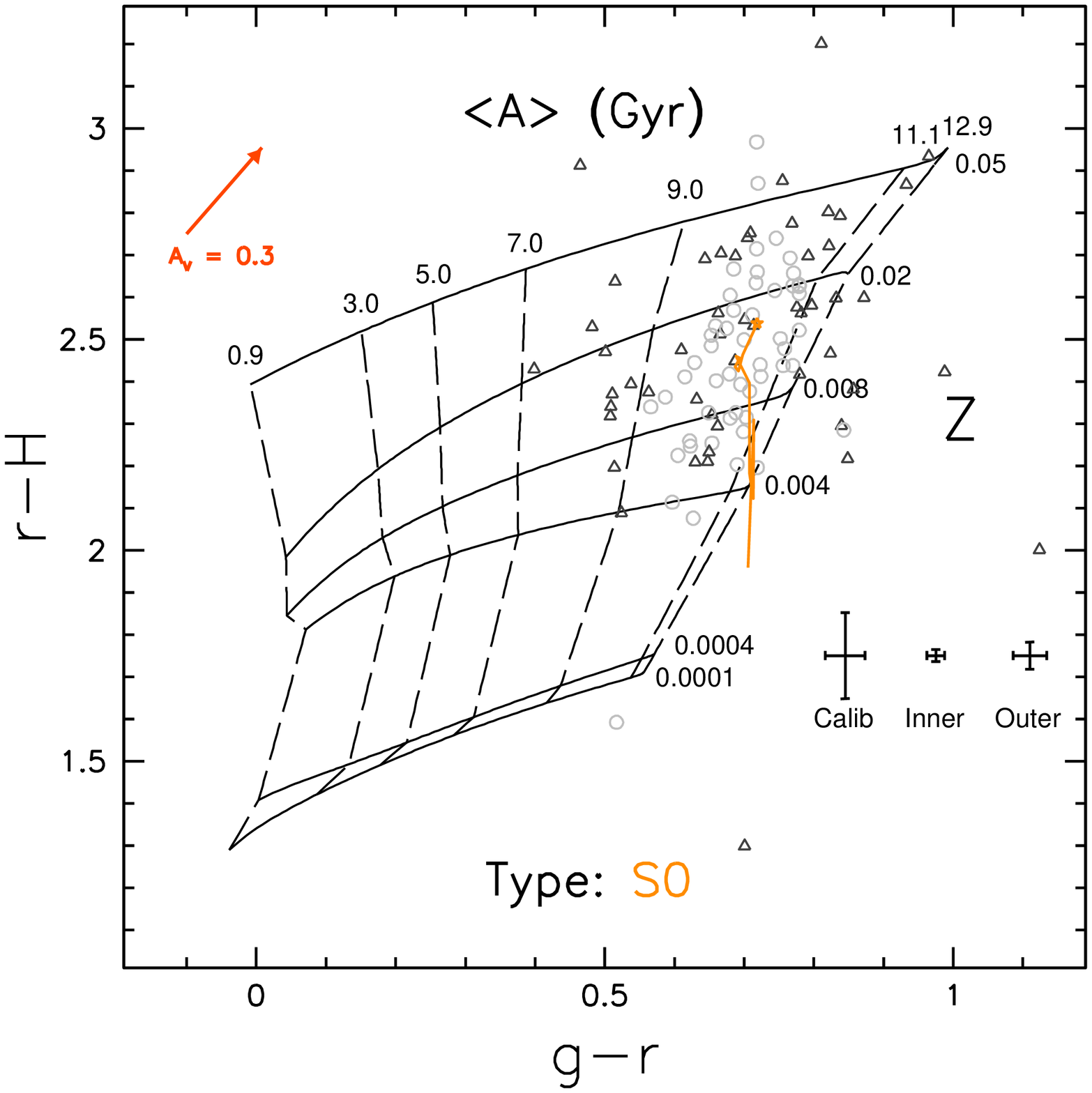} &
   \includegraphics[width=0.45\textwidth]{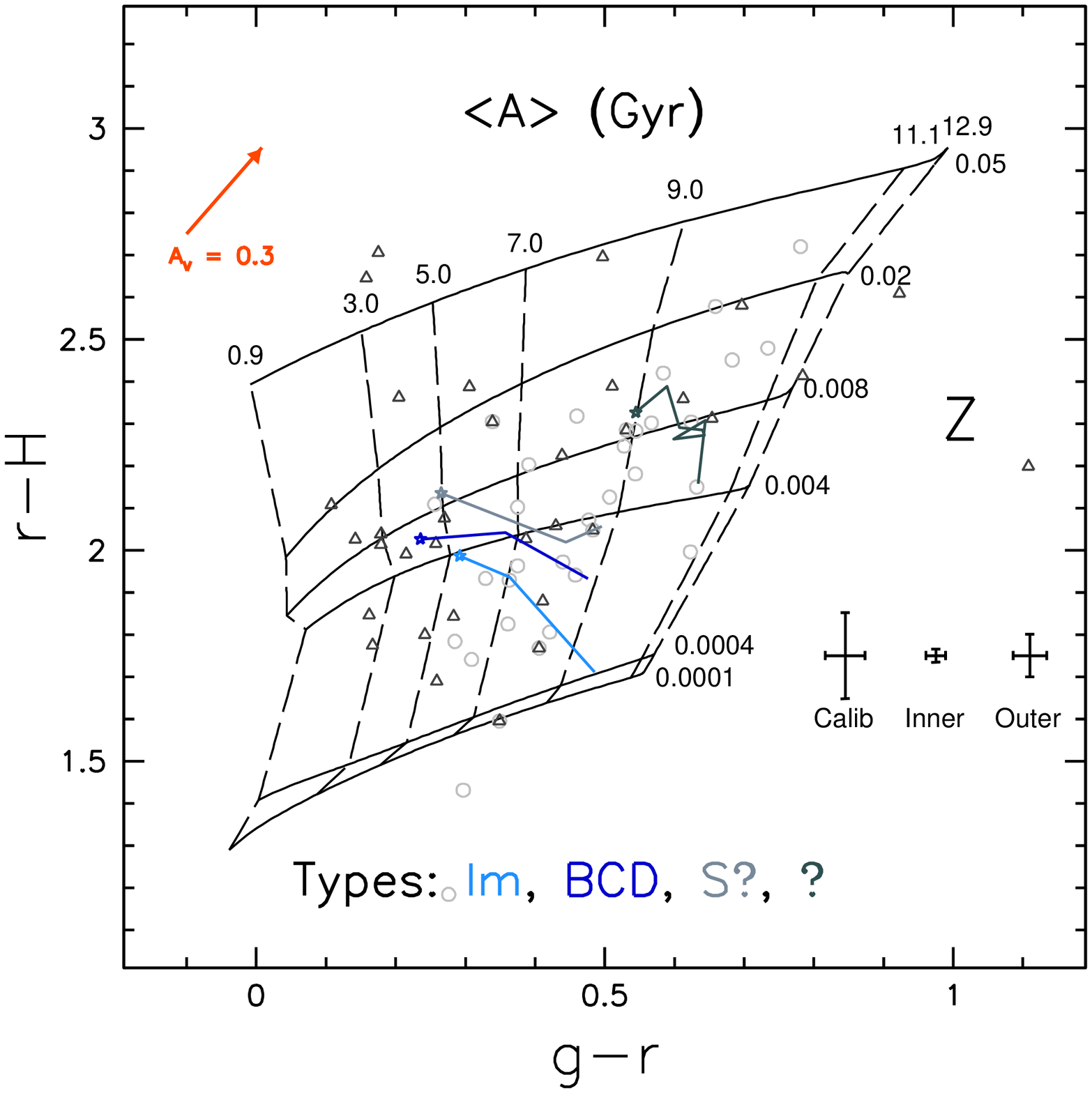}
  \end{tabular}
  \caption{Same as Fig. \ref{fig:CC-ETG} for Virgo lenticular (\emph{left}) 
and irregular (\emph{right}) galaxies.}
  \label{fig:CC-S0-Irr}
 \end{center}
\end{figure*}

\begin{figure*}
 \begin{center}
  \begin{tabular}{c c}
   \includegraphics[width=0.45\textwidth]{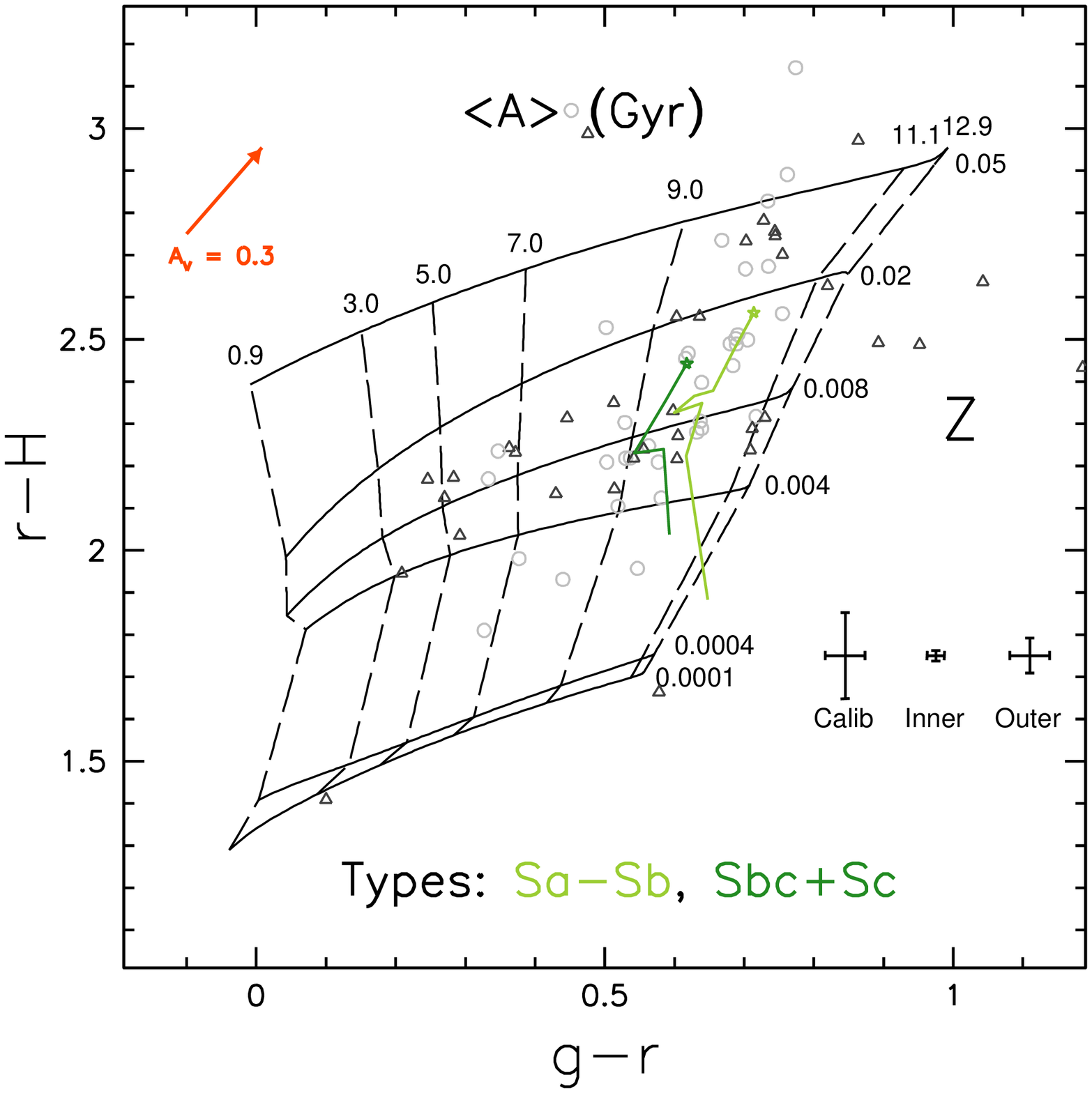} &
   \includegraphics[width=0.45\textwidth]{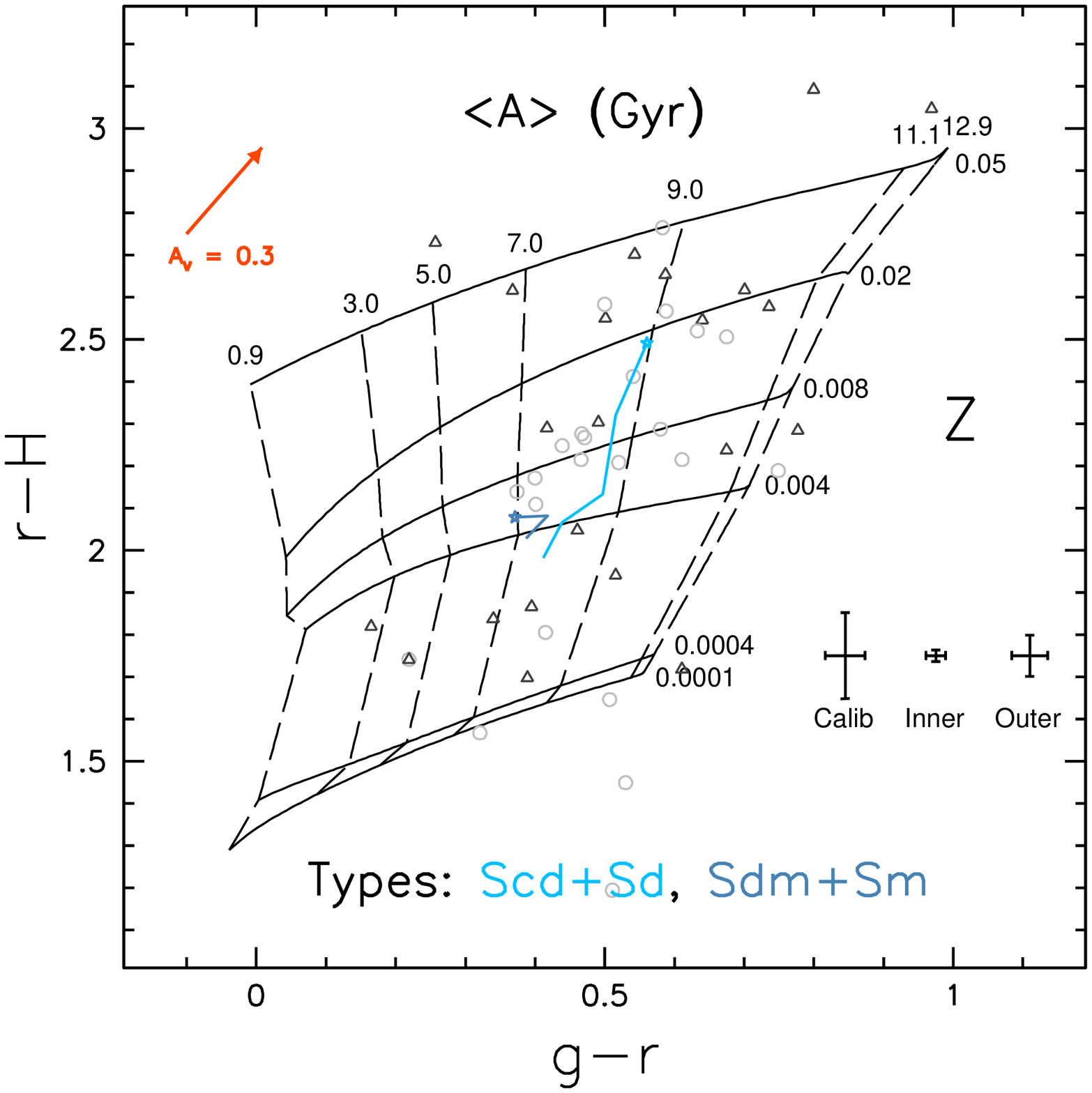}
  \end{tabular}
  \caption{Same as Fig. \ref{fig:CC-ETG} for Virgo spiral galaxies of types 
Sa$-$Sc (\emph{left}) and Scd$-$Sm (\emph{right}).}
  \label{fig:CC-Sp}
 \end{center}
\end{figure*}

\begin{figure*}
 \begin{center}
  \begin{tabular}{c c}
   \includegraphics[width=0.45\textwidth]{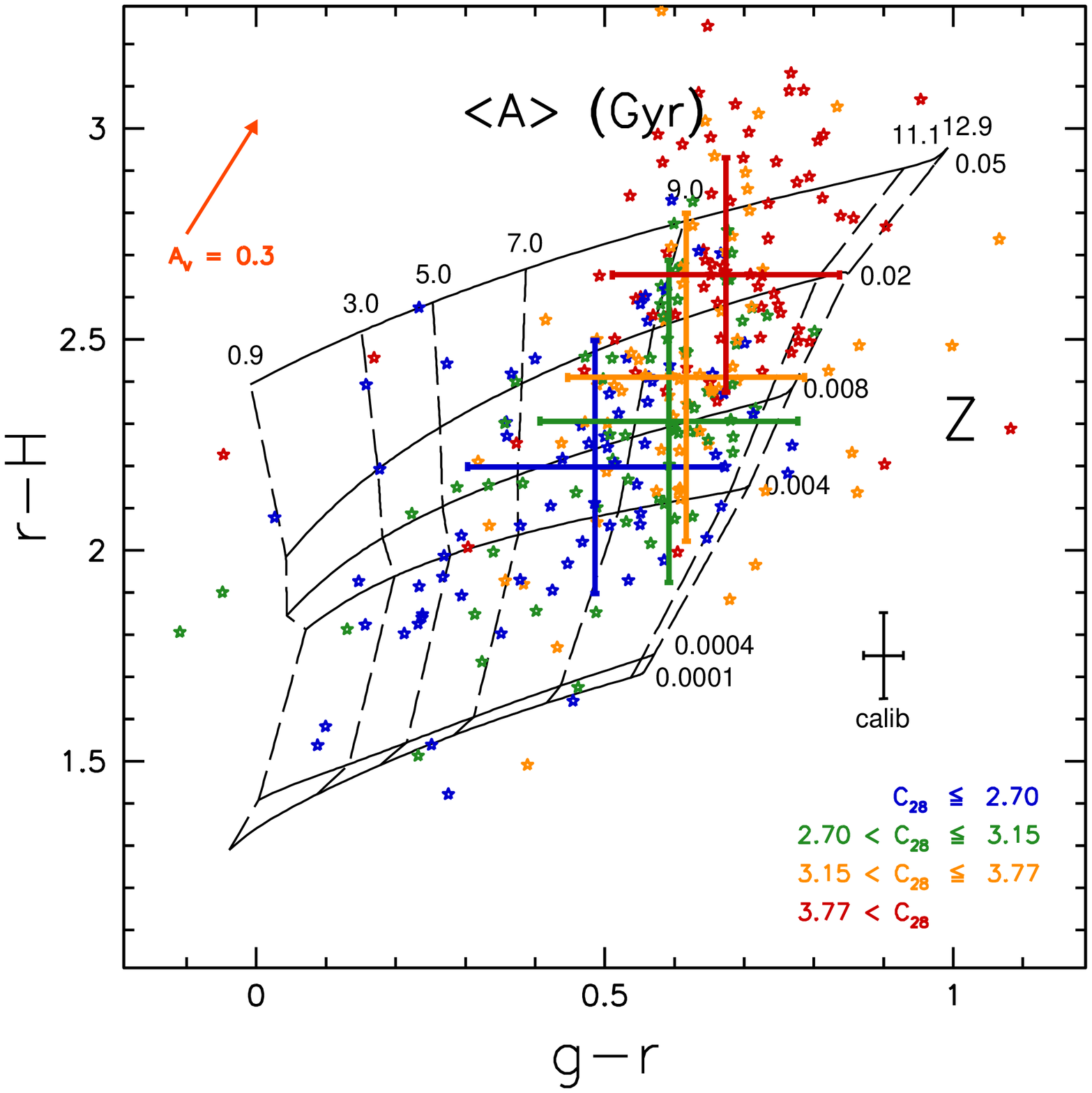} &
   \includegraphics[width=0.45\textwidth]{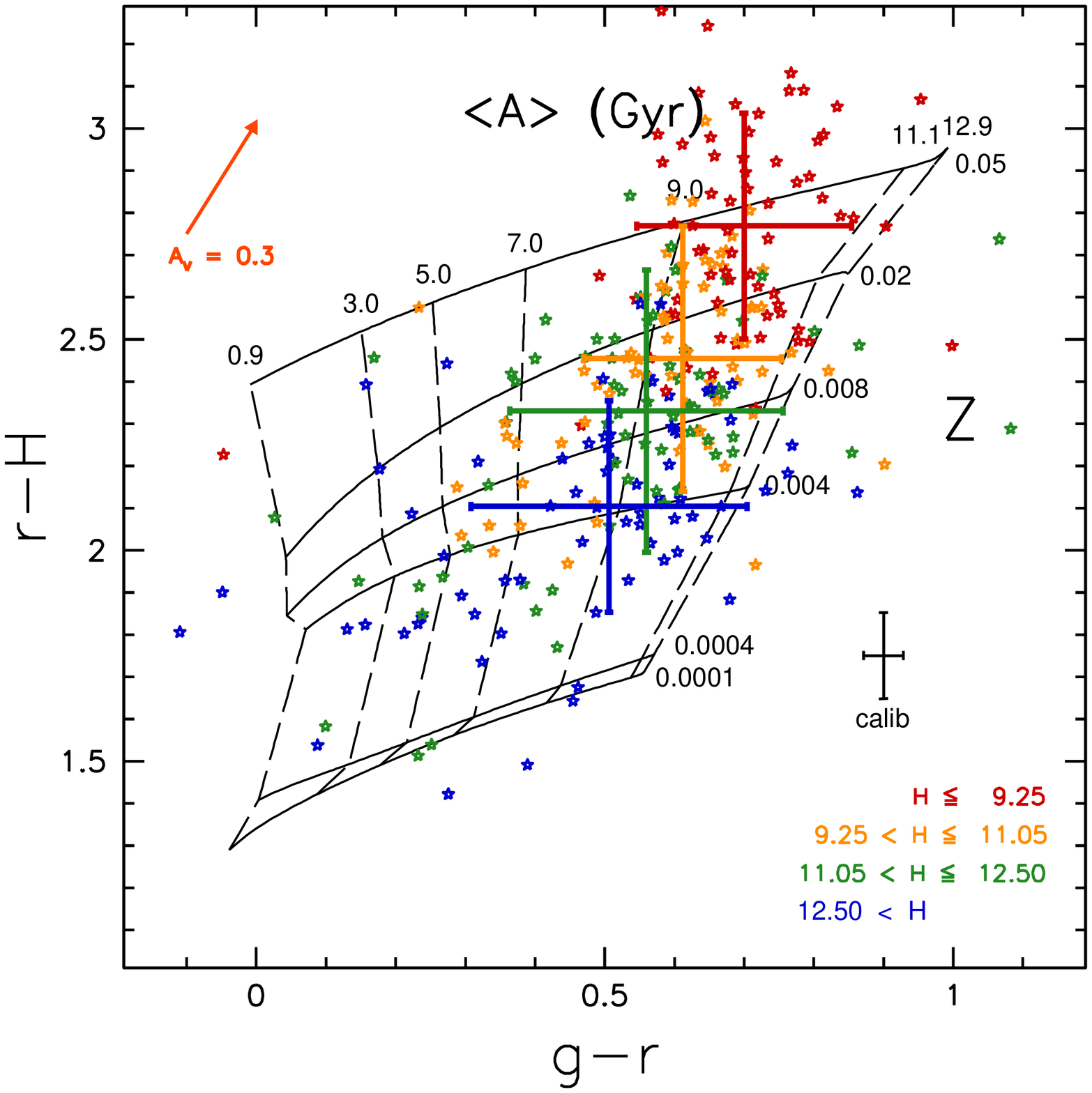} \\
   \includegraphics[width=0.45\textwidth]{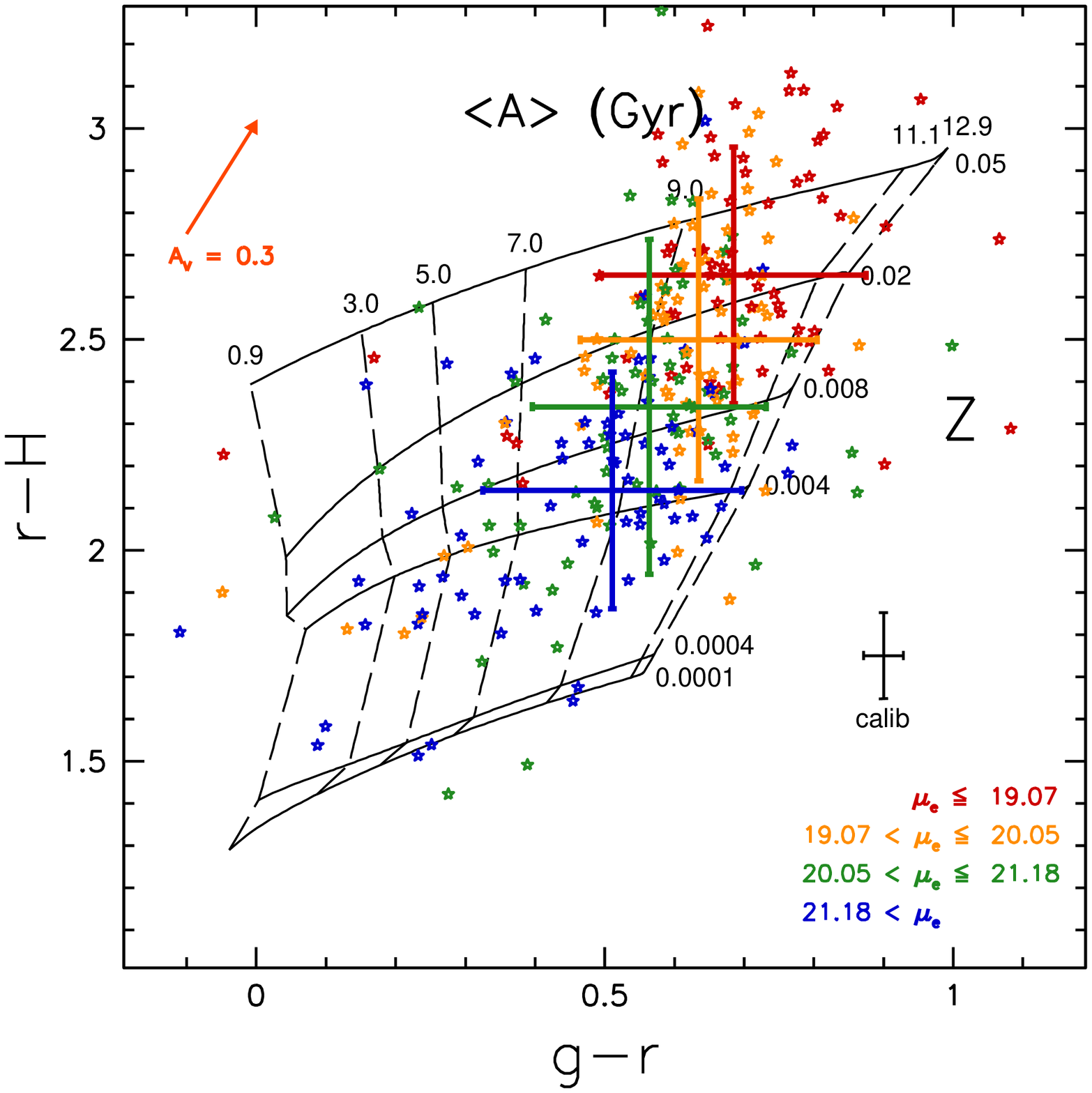} &
   \includegraphics[width=0.45\textwidth]{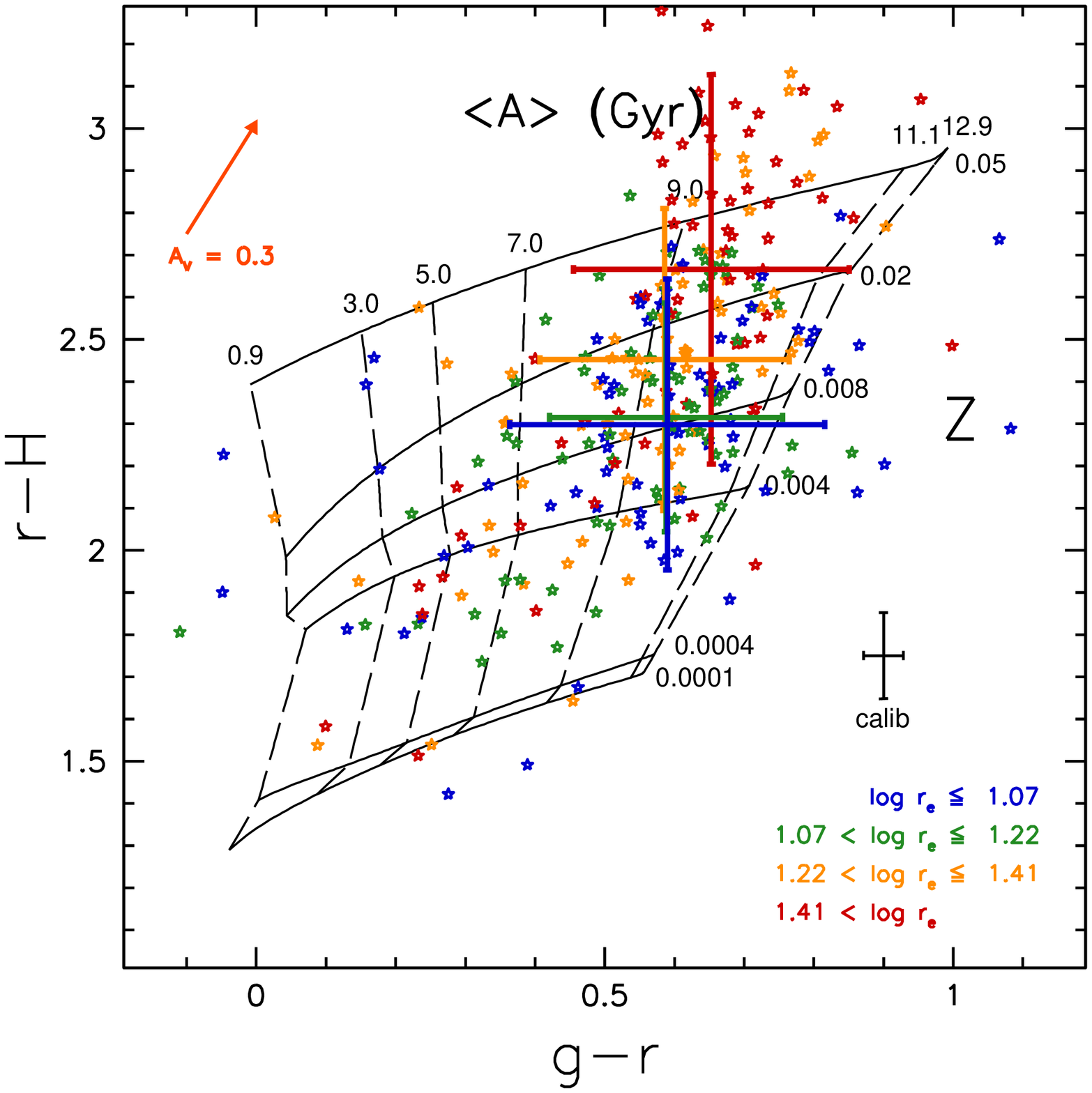}
  \end{tabular}
  \caption{(\emph{top left}) Variation in the central ($r <$ 0.5$r_e$) $r$-$H$ 
and $g$-$r$ colours with concentration $C_{28}$ for Virgo galaxies.
Galaxies have been grouped into equally-populated bins by $C_{28}$, with 
the bin ranges and point colours indicated in the lower-right corner of the 
window. The overplotted error bars show both the centroid and rms scatter of 
the colour distribution within each $C_{28}$ bin. The other plots are the same 
as at top-left, but show the variations in central $r$-$H$ and $g$-$r$ colours 
of Virgo galaxies with apparent magnitude $m_H$ (\emph{top right}), effective 
surface brightness $\mu_e$ (\emph{bottom left}), and effective radius $r_e$ 
(\emph{bottom right}).}
  \label{fig:CC-Struct}
 \end{center}
\end{figure*}

\begin{figure*}
 \begin{center}
  \begin{tabular}{c c}
   \includegraphics[width=0.45\textwidth]{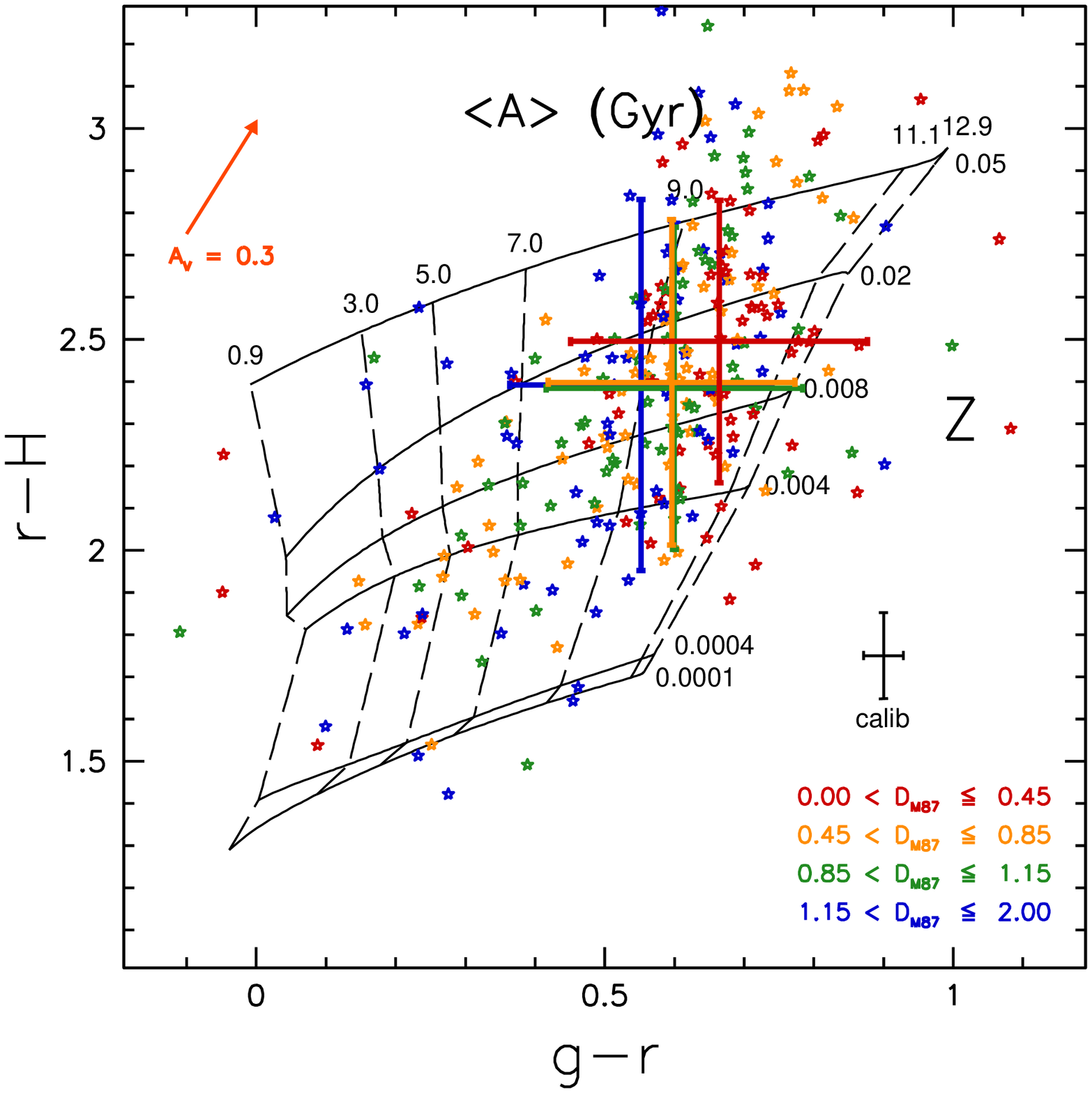} &
   \includegraphics[width=0.45\textwidth]{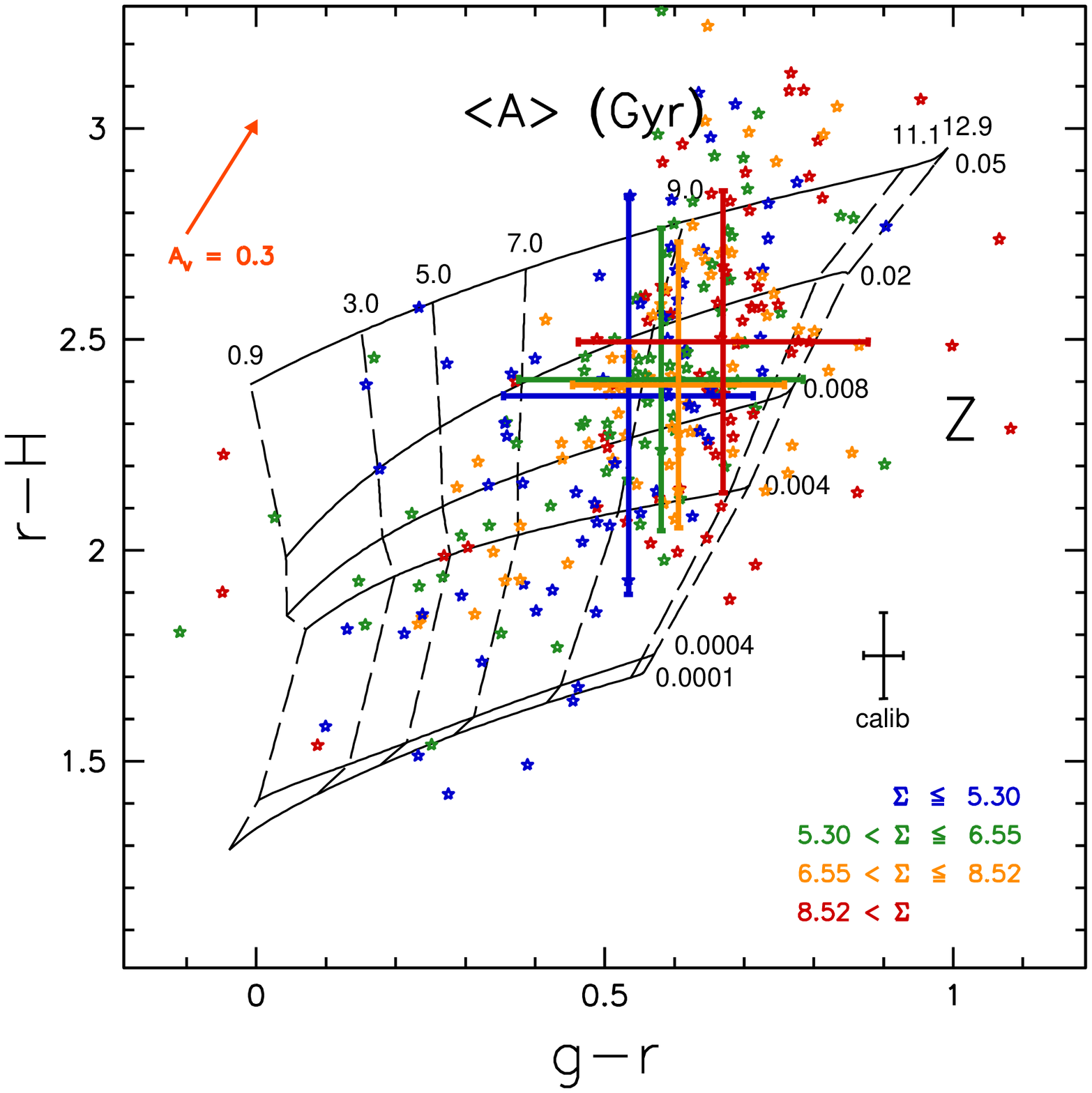} \\
  \end{tabular}
  \includegraphics[width=0.45\textwidth]{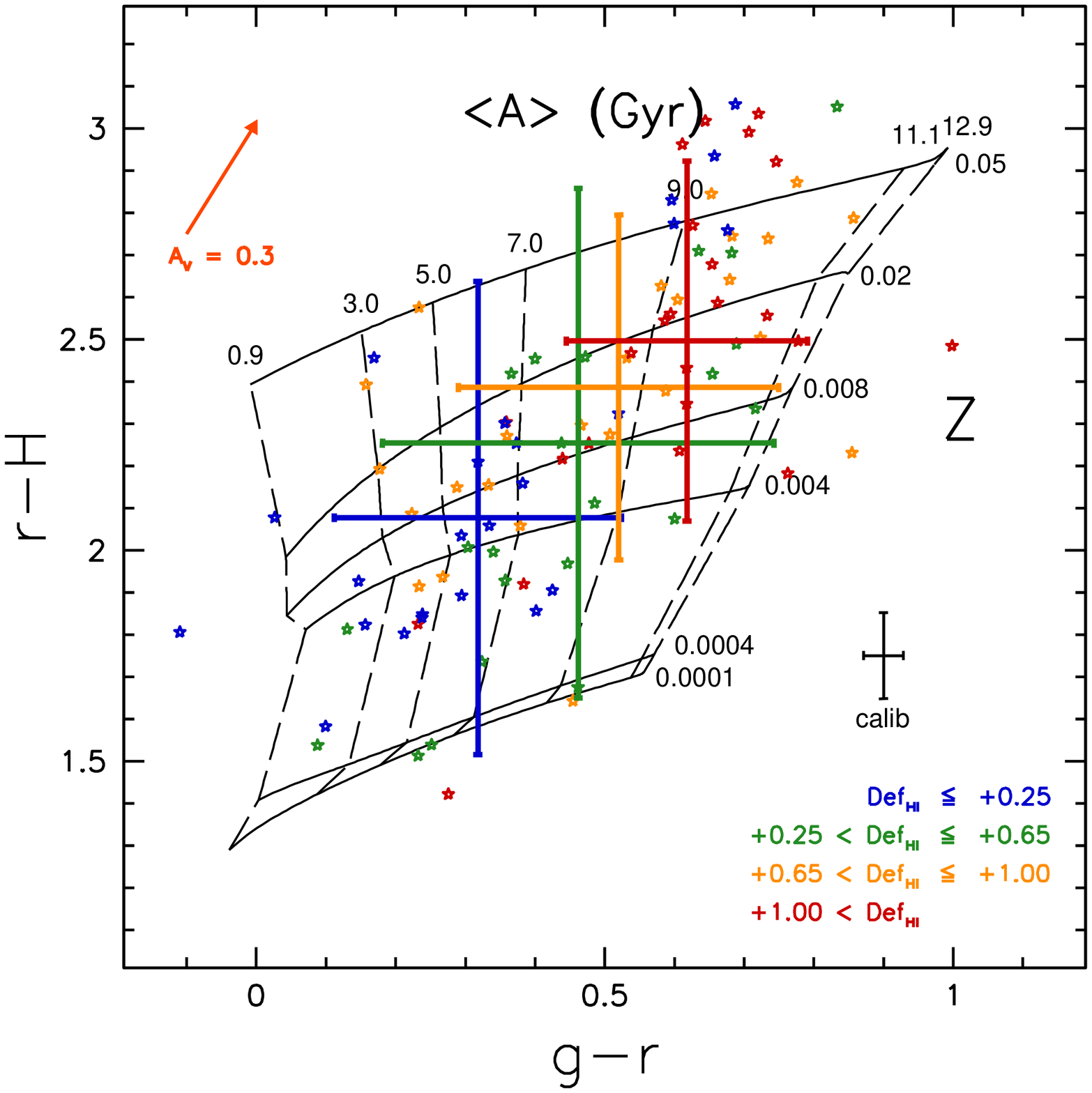}
  \caption{Same as Fig. \ref{fig:CC-Struct} but showing variations in central 
$r$-$H$ and $g$-$r$ colours of Virgo galaxies with distance from M87 $D_{M87}$ 
(\emph{top left}), local galaxian surface density $\Sigma$ (\emph{top right}) 
and neutral gas deficiency $Def_{\hi}$ (\emph{bottom}). The plot against 
$Def_{\hi}$ only includes 101 galaxies due to the limited \hi mapping 
of the Virgo cluster.}
  \label{fig:CC-Env}
 \end{center}
\end{figure*}

\begin{figure*}
 \begin{center}
  \includegraphics[width=0.9\textwidth]{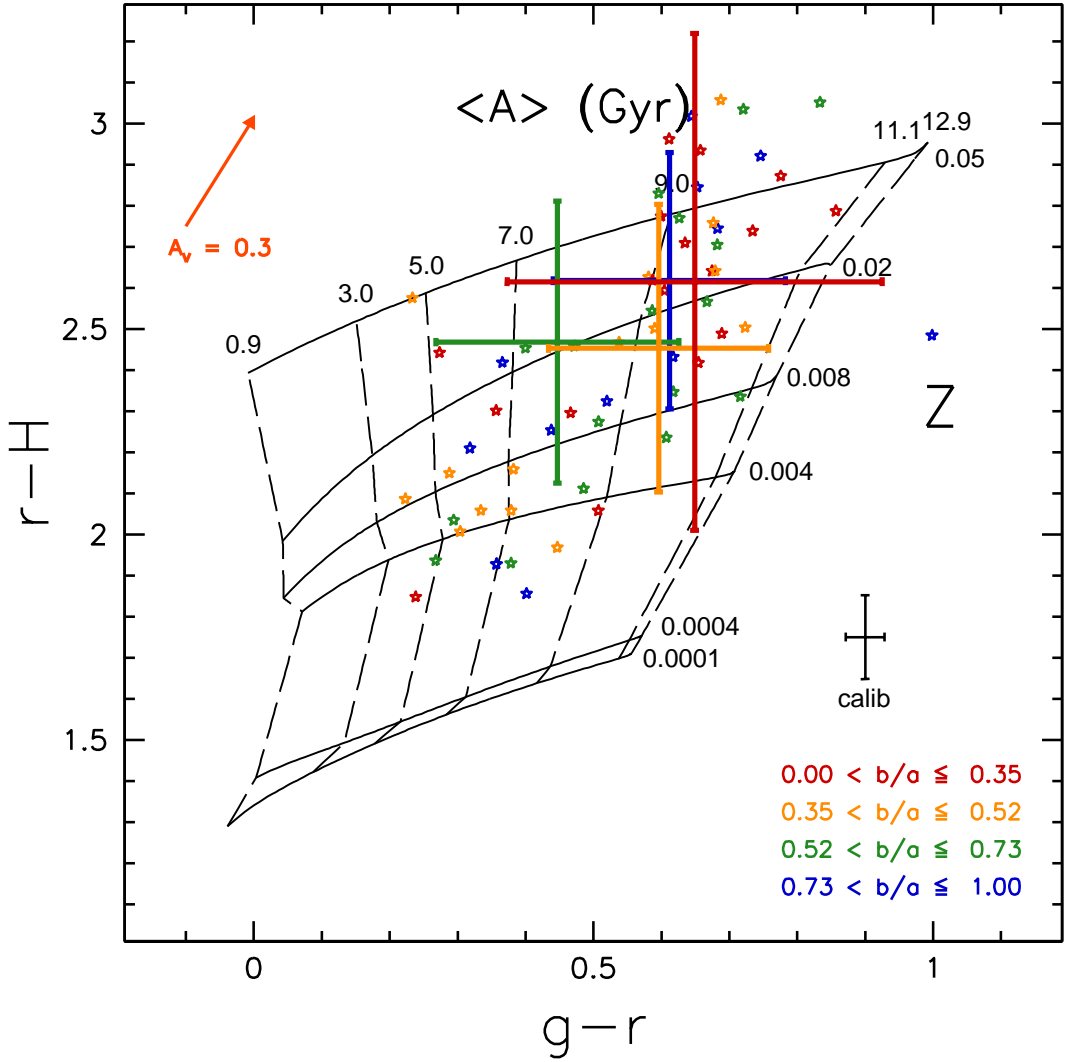}
  \caption{Same as Fig. \ref{fig:CC-Struct} but showing variations in central 
$r$-$H$ and $g$-$r$ colours of Virgo galaxies with $H$-band axis ratio, $b/a$. 
Only the colours of Virgo spiral galaxies are plotted here.}
  \label{fig:CC-ba}
 \end{center}
\end{figure*}

We now investigate trends in Virgo galaxy colours as a function of galaxy 
morphology, structure, environment and axis ratio. Beginning with morphology, 
we show colour-colour diagrams in Figures~\ref{fig:CC-ETG}-\ref{fig:CC-Sp} for 
several galaxy types against the colour predictions from a stellar population 
model (black lines) in order to diagnose the effects of mean age 
$\left\langle A\right\rangle$ ($g$-$r$; dashed) and metallicity $Z$ ($r$-$H$; 
solid) in setting the Virgo galaxy colours. The model is an exponential star 
formation history with a variable timescale $\tau$ and several fixed 
metallicities. Nominal values of both the $\left\langle A\right\rangle$ and $Z$ 
coordinates are indicated alongside the grid in each plot. Further details 
about this and other stellar population models are given in Paper II. The 
orange-red arrow represents the reddening vector for a dust screen, also 
discussed more fully in Paper II. Median values of both the calibration and sky 
errors in the M11 sample are shown in the lower-right corner of each plot, 
where the effect of the latter has been computed at both the centers and 
outskirts of Virgo galaxies. Note that both the model predictions and reddening 
vector discussed above have been included in 
Figures~\ref{fig:CC-Struct}-\ref{fig:CC-ba} as well.

Figures~\ref{fig:CC-ETG}-\ref{fig:CC-Sp} show that the different Virgo galaxy 
types have unique colour profile combinations, and thus stellar contents. The 
colour profiles of the gas-poor systems 
(Figures~\ref{fig:CC-ETG}-\ref{fig:CC-S0-Irr}) suggest that their stellar 
populations are old ($>$9 Gyr) and span a wide range in metallicity; that is, 
these colour gradients are largely explained as a metallicity effect 
\citep{Va88,Fr89,Ta00,Wu05}, while age variations make a secondary, but 
non-negligible, contribution (\citealt{LB10}; \citealt{Su10}; but see 
\citealt{BG90}). The systematic differences in the median colour profiles 
between the Virgo gas-poor giants and dwarfs (\se{CP&G}) are similarly 
explained by the stellar populations of the former being more metal-rich 
($\Delta \text{log}Z \lesssim$ -0.25), but only slightly older ($\sim$0.5 Gyr), 
than the latter. Interestingly, the dE's and E's together define the 
metallicity extrema for the entire Virgo gas-poor galaxy population, as opposed 
to the dS0's and S0's.

The Virgo irregulars (\Fig{CC-S0-Irr}) show the greatest colour range amongst 
all of the major galaxy types in this cluster such that, collectively, their 
stellar populations go from being old and metal-rich to young and metal-poor. 
This extreme colour variation extends to the Virgo irregulars as well, as their 
median colour profiles show that significant radial increases and decreases in 
age and metallicity, respectively, occur within each galaxy type. Such broad 
population gradients make it difficult to assign typical ages and metallicities 
to irregular galaxies. The offsetting effect that such radial behaviours in 
stellar populations have on colours helps explain previous reports of flat (or 
weakly positive) optical colour gradients in irregular galaxies 
\citep{KA90,Pa96,Ta05}. Moreover, the redder disks that have been found in 
dIrr's relative to those in BCDs \citep{PT96} may be understood by the former 
having older outskirts.

Age and metallicity gradients are also apparent within Virgo spirals 
(\Fig{CC-Sp}), but with the latter being more significant, on average, than the 
former. Age and metallicity variations have been ascribed previously as the 
source of the significant colour gradients in spiral galaxies 
\citep{dJ96,GdA01,LB03,Ma04}, while our claim that the implied age gradients in 
spirals are small is only matched by \cite{PB96}. In the median, the 
metallicity gradients for all Virgo spiral types appear to be uniformly 
negative, while their age gradients are more varied. This complex pattern of 
Virgo spirals' colour gradients thwarts any realistic estimate of typical ages 
and metallicities for these galaxies as well. However, the colours of Virgo 
spirals exhibit a morphological trend, also reported elsewhere 
\citep{dJ96,Ma04}, which may be explained by the earlier galaxies being older 
and more enriched than the later ones. 

\subsubsection{Structure}\label{sec:CT-Struct}
We show in \Fig{CC-Struct} the same colour-colour diagram as in \Fig{CC-ETG} 
(including the same model) but with the points differentiated according to the 
following structural parameters: $C_{28}$, $m_H$, $\mu_e$ or $r_e$. In these 
plots, we focus on galaxy centers (i.e., within an aperture of 0.5 $r_e$) to 
ease the distinction of colour trends as a function of galaxy structure. We 
also indicate with error bars the median values and rms dispersions of the 
colour distributions for all structural parameter bins.

The Virgo galaxy central colours exhibit multiple trends against structural 
parameters. The trend versus $C_{28}$ suggests that the most concentrated Virgo 
galaxies are the oldest and most metal-rich. Since $C_{28}$ measures the 
shape/concentration of a galaxy's light profile, this trend essentially 
confirms that found above versus morphology between the Virgo spheroidal and 
disk galaxies. The other trends, with $m_H$ and $\mu_e$, indicate that the more 
luminous and brighter Virgo galaxies tend to be older and more enriched than 
less luminous systems. Our detection of these trends agrees with the luminosity 
and surface brightness dependencies of galaxy colours and colour gradients 
reported in numerous studies 
\citep{Va88,Pe90a,BP94,dJ96,Ja00,Je00,TO03,Ma04,Li09} (but see \citealt{KA90} 
and \citealt{LB04} for arguments against the luminosity dependence of gas-poor 
galaxies' colours). No significant colour trend is seen against $r_e$ other 
than the most compact Virgo galaxies being older and more metal-rich than 
larger ones. This null result agrees with the weak size trends in spiral 
galaxies' colour gradients seen elsewhere \citep{MM07,Li09}, though \cite{Pa02} 
reported a disk scale length dependence in the colour gradients of field and 
group dIrr's.

\subsubsection{Environment}\label{sec:CT-Enviro}
\Fig{CC-Env} is the environmental corollary to \Fig{CC-Struct}. To investigate 
possible colour-environment trends, the M11 galaxies have been assigned in each 
of these figures to one of four equally-populated bins in the following 
parameters: $D_{M87}$, $\Sigma$, or $Def_{\hi}$. We find from these plots that 
the median colours of the different $D_{M87}$ and $\Sigma$ bins all roughly 
coincide, suggesting that cluster-centric location of, or local density about, 
a Virgo galaxy has little influence on its stellar content. This result is 
matched by \cite{Ko05} and \cite{MM07} who found no change in the colours or 
colour gradients of gas-rich galaxies with local density. The contradictory 
claim by \citep{Pa02} about a density dependence of galaxy colours is diffused 
by considering that those authors used a sample which spanned a wider density 
baseline (field$\rightarrow$cluster) than ours.

Amongst Virgo gas-rich galaxies, we find a clean trend of colours with 
$Def_{\hi}$, such that more deficient systems are older and more metal-rich than 
less deficient ones. However, the comparable $r$-$H$ colours found amongst 
intermediate-$Def_{\hi}$ galaxies suggest that age effects dominate over those 
of metallicity in creating the observed trend. To our knowledge, this is the 
first report of a gas deficiency-colour correlation for Virgo galaxies. 
However, a latent correlation between the $Def_{\hi}$ measurements for our 
galaxies and their morphology, with a Pearson correlation coefficient of -0.68, 
makes it unclear whether the colour-$Def_{\hi}$ trend is fundamental or not.

\subsubsection{Dust}\label{sec:CT-Dust}
It would also be prudent to address the effects of dust on the colour 
properties of Virgo cluster galaxies. The presence of dust in star-forming 
systems is of concern in stellar population studies since reddened broadband 
colours may be erroneously attributed to their stellar content. For instance, 
the colour gradients of a galaxy with a smoothly-varying dust distribution may 
be taken as due to combined negative age and metallicity gradients within it, a 
consequence of the three-fold age-metallicity-dust degeneracy. Even face-on 
systems may be susceptible to such effects \citep[M04; Paper II]{Pe99}. 
Moreover, \cite{GdJ95} have shown that the shallow optical colour gradients of 
gas-poor giants may be explained by a diffuse dust component, while \cite{Ga09} 
have found evidence for severe (but localized) dust attenuation in the 
optical-NIR profiles of nearby spirals.

In \Fig{CC-ba}, we assess the dust reddening of the central colours of Virgo 
spirals in the $r$-$H$ versus $g$-$r$ plane, where individual galaxies are
identified according to their NIR axis ratio, $b/a$. In so doing, we find large 
dispersions in the colour distributions for edge-on systems, which suggests 
that some of them likely suffer from significant reddening. The colour 
centroids of the highest and lowest inclination spirals are nearly identical, 
however, which is inconsistent with dust being the main driver of these 
galaxies' central colours. That is, in a statistical sense, the effect of dust 
on our galaxy colours may be neglected. This conclusion finds ample support in 
the literature \citep{dJ96,PB96,GdA01,Ma04}, but does not preclude a possible 
contribution of dust to the colour gradients of individual systems 
\citep[e.g.,][]{Pe95,LB03}. Indeed, we find dramatic colour variations between 
adjacent radial bins in a few Virgo spirals, indicative of significant 
reddening at those (redward) locations (i.e., those colour variations would 
otherwise imply radical changes in local age and metallicity), but the 
incidence of such bins is small.


\section{CONCLUSIONS}\label{sec:Concs}

By combining high-quality optical ($gri$) and near-infrared ($H$) imaging for 
$\sim$300 Virgo cluster galaxies, we have studied the resolved broadband 
colours of a representative galaxy sample spanning the full range of structural 
parameters. Below is a summary of our main findings about Virgo galaxies' 
colour gradients and trends in their central colours with structural and 
environmental parameters:
\begin{itemize}
 \item In the median, the colour gradients of Virgo galaxies systematically 
   decrease with morphological type, such that the gradients of giant 
   early-types are nearly flat, while those of later types are strongly 
   negative (i.e., red centers to blue outskirts). The colour gradients of 
   early-type dwarfs are more negative than those for the giants;
\item The colour gradients of all Virgo galaxies principally arise from stellar 
   metallicity gradients, while stellar age gradients contribute only to the 
   colours of Virgo irregulars;
 \item The central colours of Virgo galaxies correlate with concentration, 
   luminosity, and surface brightness, such that galaxies become redder with an 
   increase in each of these parameters;
 \item While the central colours of Virgo galaxies do not depend on 
   cluster-centric distance or local galaxy density, those of Virgo gas-rich 
   galaxies correlate strongly with their neutral gas deficiency;
 \item The central colours of Virgo spiral galaxies are uncorrelated with axial 
   ratio, thus limiting the role of dust in setting their colours. 
\end{itemize}
While the above results reflect on the stellar content of Virgo galaxies and 
their structural and environmental dependencies, we can explicitly apply 
stellar population models to galaxy colours. This is what we do in Paper II
as a way to constrain formation scenarios for the major galaxy types in the 
Virgo cluster.

\bigskip
This research has made use of the NASA/IPAC extragalactic and GOLDMine 
databases. Funding for the SDSS has been provided by the Alfred P. Sloan 
Foundation, the Participating Institutions, the National Science Foundation, 
the US Department of Energy, the National Aeronautics and Space Administration, 
the Japanese Monbukagakusho, the Max Planck Society, and the Higher Education 
Funding Council for England. The SDSS Web Site is http://www.sdss.org/. The 
SDSS is managed by the Astrophysical Research Consortium for the Participating 
Institutions. The Participating Institutions are the American Museum of Natural 
History, Astrophysical Institute Potsdam, University of Basel, Cambridge 
University, Case Western Reserve University, University of Chicago, Drexel 
University, Fermilab, the Institute for Advanced Study, the Japan Participation 
Group, The Johns Hopkins University, the Joint Institute for Nuclear 
Astrophysics, the Kavli Institute for Particle Astrophysics and Cosmology, the 
Korean Scientist Group, the Chinese Academy of Sciences (LAMOST), Los Alamos 
National Laboratory, the Max-Planck-Institute for Astronomy, the 
Max-Planck-Institute for Astrophysics, New Mexico State University, The Ohio 
State University, University of Pittsburgh, University of Portsmouth, Princeton 
University, the United States Naval Observatory, and the University of 
Washington. J.R. and S.C. acknowledge financial support from the National 
Science and Engineering Council of Canada through a postgraduate scholarship 
and a Discovery Grant, respectively. We also thank Brent Tully for his help
in the acquisition of $H$-band imaging of Virgo cluster galaxies (as reported
in M11). 


\clearpage


\begin{deluxetable}{cccccc}
 \tabletypesize{\scriptsize}  
 \tablewidth{0pc}  
 \tablecaption{Median $g$-$H$ colours and colour gradients of Virgo galaxies}
 \tablehead{
  \colhead{Morphology} &
  \colhead{} &
  \colhead{$\dfrac{\text{d}(g\text{-}H)}{\text{d}r}^{\text{a}}$} &
  \colhead{} &
  \colhead{$(g\text{-}H)_0$} &
  \colhead{$N$} \\ \\ [-5pt]
  \cline{2-4} \\ [-5pt]
  \colhead{} &
  \colhead{$r/r_e$ (mag $r_e^{-1}$)} &
  \colhead{$r$ (mag kpc$^{-1}$)} &
  \colhead{log $r$ (mag dex$^{-1}$)} &
  \colhead{} &
  \colhead{} \\ [2pt]
  \colhead{(1)} &
  \colhead{(2)} &
  \colhead{(3)} &
  \colhead{(4)} &
  \colhead{(5)} &
  \colhead{(6)}
 }
 \startdata
 dS0     & -0.126 $\pm$ 0.199 & -0.080 $\pm$ 0.165 & -0.198 $\pm$ 0.290 & 3.000 $\pm$ 0.220 & 19 \cr
 dE      & -0.177 $\pm$ 0.208 & -0.167 $\pm$ 0.196 & -0.309 $\pm$ 0.377 & 2.843 $\pm$ 0.184 & 51 \cr
 E       & -0.084 $\pm$ 0.199 & -0.075 $\pm$ 0.339 & -0.267 $\pm$ 0.572 & 3.253 $\pm$ 0.353 & 32 \cr
 S0      & -0.046 $\pm$ 0.212 & -0.028 $\pm$ 0.152 & -0.115 $\pm$ 0.310 & 3.283 $\pm$ 0.280 & 53 \cr
 Sa$-$Sb & -0.138 $\pm$ 0.314 & -0.057 $\pm$ 0.080 & -0.322 $\pm$ 0.424 & 3.274 $\pm$ 0.578 & 24 \cr
 Sbc+Sc  & -0.190 $\pm$ 0.324 & -0.066 $\pm$ 0.101 & -0.428 $\pm$ 0.407 & 3.176 $\pm$ 0.663 & 15 \cr
 Scd+Sd  & -0.259 $\pm$ 0.412 & -0.106 $\pm$ 0.180 & -0.371 $\pm$ 0.652 & 3.037 $\pm$ 0.588 & 17 \cr
 Sdm+Sm  & -0.017 $\pm$ 0.467 & -0.009 $\pm$ 0.231 & -0.072 $\pm$ 0.616 & 2.459 $\pm$ 0.420 &  9 \cr
 Im      & -0.181 $\pm$ 0.391 & -0.076 $\pm$ 0.278 & -0.228 $\pm$ 0.662 & 2.302 $\pm$ 0.499 & 13 \cr
 BCD     & +0.051 $\pm$ 0.233 & +0.071 $\pm$ 0.321 & +0.197 $\pm$ 0.403 & 2.355 $\pm$ 0.417 &  8 \cr
 S?      & +0.141 $\pm$ 0.400 & +0.108 $\pm$ 0.213 & +0.158 $\pm$ 0.468 & 2.415 $\pm$ 0.466 &  9 \cr
 ?       & -0.137 $\pm$ 0.178 & -0.077 $\pm$ 0.266 & -0.246 $\pm$ 0.479 & 2.850 $\pm$ 0.422 &  8 \\
 \enddata
 \tablenotetext{a}{The median colour gradients for Virgo galaxies were computed 
for three choices of radial coordinate: effective radii 
(\textit{left}), kiloparsec (\textit{center}), and logarithmic 
(\textit{right}).}
 \label{tbl:Cgrd}
\end{deluxetable}


\clearpage

\clearpage

\clearpage

\clearpage

\clearpage

\clearpage

\clearpage

\clearpage

\clearpage

\clearpage

\clearpage

\clearpage


\begin{thebibliography}{99}
\bibitem[Abraham et al.(1999)]{Ab99} Abraham, R.~G., Ellis, R.~S., Fabian, 
A.~C., Tanvir, N.~R., \& Glazebrook, K.\ 1999, \mnras, 303, 641
\bibitem[Adelman-McCarthy et al.(2007)]{Ad07} Adelman-McCarthy, J.~K., et al.\ 
2007, \apjs, 172, 634
\bibitem[Adelman-McCarthy et al.(2008)]{Ad08} Adelman-McCarthy, J.~K., et al.\ 
2008, \apjs, 175, 297
\bibitem[Balcells \& Peletier(1994)]{BP94} Balcells, M., \& Peletier, R.~F.\ 
1994, \aj, 107, 135
\bibitem[Barazza et al.(2001)]{Ba01} Barazza, F.~D., Binggeli, B., \& Prugniel, 
P.\ 2001, \aap, 373, 12
\bibitem[Barazza et al.(2003)]{Ba03} Barazza, F.~D., Binggeli, B., \& Jerjen, 
H.\ 2003, \aap, 407, 121
\bibitem[Bell \& de Jong(2000)]{BdJ00} Bell, E.~F., \& de Jong, R.~S.\ 2000, 
\mnras, 312, 497
\bibitem[Bergvall et al.(1999)]{Be99} Bergvall, N., R{\"o}nnback, J., Masegosa, 
J., {\&Ouml}stlin, G.\ 1999, \aap, 341, 697
\bibitem[Binggeli et al.(1985)]{Bi85} Binggeli, B., Sandage, A., \& Tammann, 
G.~A.\ 1985, \aj, 90, 1681
\bibitem[Boselli et al.(2008)]{Bo08} Boselli, A., Boissier, S., Cortese, L., \& 
Gavazzi, G.\ 2008, \apj, 674, 742
\bibitem[Bothun \& Gregg(1990)]{BG90} Bothun, G.~D., \& Gregg, M.~D.\ 1990, 
\apj, 350, 73
\bibitem[Bremnes et al.(1998)]{Br98} Bremnes, T., Binggeli, B., \& Prugniel, 
P.\ 1998, \aaps, 129, 313
\bibitem[Bremnes et al.(1999)]{Br99} Bremnes, T., Binggeli, B., \& Prugniel, 
P.\ 1999, \aaps, 137, 337
\bibitem[Bremnes et al.(2000)]{Br00} Bremnes, T., Binggeli, B., \& Prugniel, 
P.\ 2000, \aaps, 141, 211
\bibitem[Burstein et al.(1984)]{Bu84} Burstein, D., Faber, S.~M., Gaskell, 
C.~M., \& Krumm, N.\ 1984, \apj, 287, 586
\bibitem[Cardiel et al.(2003)]{Ca03} Cardiel, N., Gorgas, J., 
S{\'a}nchez-Bl{\'a}zquez, P., Cenarro, A.~J., Pedraz, S., Bruzual, G., \& 
Klement, J.\ 2003, \aap, 409, 511
\bibitem[Courteau et al.(1996)]{Co96} Courteau, S., de Jong, R.~S., \& Broeils, 
A.~H.\ 1996, \apjl, 457, L73
\bibitem[de Jong(1996)]{dJ96} de Jong, R.~S.\ 1996, \aap, 313, 377
\bibitem[Dressler(1980)]{Dr80} Dressler, A.\ 1980, \apj, 236, 351
\bibitem[Faber(1973)]{Fa73} Faber, S.~M.\ 1973, \apj, 179, 731
\bibitem[Ferreras et al.(2005)]{Fe05} Ferreras, I., Lisker, T., Carollo, C.~M., 
Lilly, S.~J., \& Mobasher, B.\ 2005, \apj, 635, 243
\bibitem[Ferreras et al.(2009)]{Fe09} Ferreras, I., Lisker, T., Pasquali, A., 
\& Kaviraj, S.\ 2009, \mnras, 395, 554
\bibitem[Franx et al.(1989)]{Fr89} Franx, M., Illingworth, G., \& Heckman, T.\ 
1989, \aj, 98, 538
\bibitem[Gadotti \& dos Anjos(2001)]{GdA01} Gadotti, D.~A., \& dos Anjos, S.\ 
2001, \aj, 122, 1298
\bibitem[Gallagher \& Hunter(1986)]{GH86} Gallagher, J.~S., III, \& Hunter, 
D.~A.\ 1986, \aj, 92, 557
\bibitem[Gallazzi et al.(2005)]{Gal05} Gallazzi, A., Charlot, S., Brinchmann, 
J., White, S.~D.~M., \& Tremonti, C.~A.\ 2005, \mnras, 362, 41
\bibitem[Ganda et al.(2009)]{Ga09} Ganda, K., Peletier, R.~F., Balcells, M., \& 
Falc{\'o}n-Barroso, J.\ 2009, \mnras, 395, 1669
\bibitem[Gavazzi et al.(2002)]{Ga02} Gavazzi, G., Bonfanti, C., Sanvito, G., 
Boselli, A., \& Scodeggio, M.\ 2002, \apj, 576, 135
\bibitem[Gavazzi et al.(2003)]{Ga03} Gavazzi, G., Boselli, A., Donati, A., 
Franzetti, P., \& Scodeggio, M.\ 2003, \aap, 400, 451
\bibitem[Gavazzi et al.(2005)]{Gav05} Gavazzi, G., Boselli, A., van Driel, W., 
\& O'Neil, K.\ 2005, \aap, 429, 439
\bibitem[Gil de Paz \& Madore(2005)]{GdPM05} Gil de Paz, A., \& Madore, B.~F.\ 
2005, \apjs, 156, 345
\bibitem[Gonzalez-Perez et al.(2011)]{GP10} Gonzalez-Perez, V., Castander, 
F.~J., \& Kauffmann, G.\ 2011, \mnras, 411, 1151
\bibitem[Goudfrooij et al.(1994)]{Go94} Goudfrooij, P., Hansen, L., Jorgensen, 
H.~E., Norgaard-Nielsen, H.~U., de Jong, T., \& van den Hoek, L.~B.\ 1994, 
\aaps, 104, 179
\bibitem[Goudfrooij \& de Jong(1995)]{GdJ95} Goudfrooij, P., \& de Jong, T.\ 
1995, \aap, 298, 784
\bibitem[Grant et al.(2005)]{Gr05} Grant, N.~I., Kuipers, J.~A., \& Phillipps, 
S.\ 2005, \mnras, 363, 1019
\bibitem[Hunter \& Elmegreen(2006)]{HE06} Hunter, D.~A., \& Elmegreen, B.~G.\ 
2006, \apjs, 162, 49
\bibitem[James(1994)]{Ja94} James, P.~A.\ 1994, \mnras, 269, 176
\bibitem[James et al.(2006)]{Ja06} James, P.~A., Salaris, M., Davies, J.~I., 
Phillipps, S., \& Cassisi, S.\ 2006, \mnras, 367, 339
\bibitem[Jansen et al.(2000)]{Ja00} Jansen, R.~A., Franx, M., Fabricant, D., \& 
Caldwell, N.\ 2000, \apjs, 126, 271
\bibitem[Jerjen et al.(2000)]{Je00} Jerjen, H., Binggeli, B., \& Freeman, 
K.~C.\ 2000, \aj, 119, 593
\bibitem[Kim \& Ann(1990)]{KA90} Kim, K.~O., \& Ann, H.~B.\ 1990, Journal of 
Korean Astronomical Society, 23, 43
\bibitem[Ko \& Im(2005)]{KI05} Ko, J., \& Im, M.\ 2005, Journal of Korean Astronomical Society, 38, 149
\bibitem[Kobayashi \& Arimoto(1999)]{KA99} Kobayashi, C., \& Arimoto, N.\ 1999, 
\apj, 527, 573
\bibitem[Koo et al.(2005)]{Ko05} Koo, D.~C., et al.\ 2005, \apjs, 157, 175
\bibitem[Koopmann \& Kenney(2004)]{KK04} Koopmann, R.~A., \& Kenney, J.~D.~P.\ 
2004, \apj, 613, 866
\bibitem[La Barbera et al.(2002)]{LB02} La Barbera, F., Busarello, G., 
Merluzzi, P., Massarotti, M., \& Capaccioli, M.\ 2002, \apj, 571, 790
\bibitem[La Barbera et al.(2003)]{LB03} La Barbera, F., Busarello, G., 
Massarotti, M., Merluzzi, P., \& Mercurio, A.\ 2003, \aap, 409, 21
\bibitem[La Barbera et al.(2004)]{LB04} La Barbera, F., Merluzzi, P., 
Busarello, G., Massarotti, M., \& Mercurio, A.\ 2004, \aap, 425, 797
\bibitem[La Barbera et al.(2005)]{LB05} La Barbera, F., de Carvalho, R.~R., 
Gal, R.~R., Busarello, G., Merluzzi, P., Capaccioli, M., \& Djorgovski, S.~G.\ 
2005, \apjl, 626, L19
\bibitem[La Barbera \& de Carvalho(2009)]{LBdC09} La Barbera, F., \& de 
Carvalho, R.~R.\ 2009, \apjl, 699, L76
\bibitem[La Barbera et al.(2010)]{LB10} La Barbera, F., de Carvalho, R.~R., de 
la Rosa, I.~G., Gal, R.~R., Swindle, R., \& Lopes, P.~A.~A.\ 2010, 
arXiv:1006.4056
\bibitem[Liu et al.(2009)]{Li09} Liu, C.-Z., Shen, S.-Y., Shao, Z.-Y., Chang, 
R.-X., Hou, J.-L., Yin, J., \& Yang, D.-W.\ 2009, Research in Astronomy and 
Astrophysics, 9, 1119
\bibitem[MacArthur et al.(2004)]{Ma04} MacArthur, L.~A., Courteau, S., Bell, 
E., \& Holtzman, J.~A.\ 2004, \apjs, 152, 175
\bibitem[MacArthur(2005)]{Ma05} MacArthur, L.~A.\ 2005, \apj, 623, 795
\bibitem[MacArthur et al.(2009)]{Ma09} MacArthur, L.~A., Gonz{\'a}lez, J.~J., 
\& Courteau, S.\ 2009, \mnras, 395, 28
\bibitem[McDonald et al.(2009)]{Mc09} McDonald, M., Courteau, S., \& Tully, 
R.~B.\ 2009, \mnras, 394, 2022
\bibitem[McDonald et al.(2011)]{Mc11} McDonald, M., Courteau, S., Tully, R.~B., 
\& Roediger, J.~C.\ 2011, \mnras, in press
\bibitem[McLaughlin(1999)]{Mc99} McLaughlin, D.~E.\ 1999, \apjl, 512, L9
\bibitem[Mei et al.(2007)]{Me07} Mei, S., et al.\ 2007, \apj, 655, 144
\bibitem[Menanteau et al.(2004)]{Me04} Menanteau, F., et al.\ 2004, \apj, 612, 
202
\bibitem[Michard(1999)]{Mi99} Michard, R.\ 1999, \aaps, 137, 245
\bibitem[Michard(2000)]{Mi00} Michard, R.\ 2000, \aap, 360, 85
\bibitem[Moriondo et al.(2001)]{Mo01} Moriondo, G., et al.\ 2001, \aap, 370, 881
\bibitem[Moth \& Elston(2002)]{ME02} Moth, P., \& Elston, R.~J.\ 2002, \aj, 
124, 1886
\bibitem[Mu{\~n}oz-Mateos et al.(2007)]{MM07} Mu{\~n}oz-Mateos, J.~C., Gil de 
Paz, A., Boissier, S., Zamorano, J., Jarrett, T., Gallego, J., \& Madore, 
B.~F.\ 2007, \apj, 658, 1006
\bibitem[Papaderos et al.(1996)]{Pa96} Papaderos, P., Loose, H.-H., Thuan, 
T.~X., \& Fricke, K.~J.\ 1996, \aaps, 120, 207
\bibitem[Parodi et al.(2002)]{Pa02} Parodi, B.~R., Barazza, F.~D., \& Binggeli, 
B.\ 2002, \aap, 388, 29
\bibitem[Patterson \& Thuan(1996)]{PT96} Patterson, R.~J., \& Thuan, T.~X.\ 
1996, \apjs, 107, 103
\bibitem[Peletier \& Valentijn(1989)]{PV89} Peletier, R.~F., \& Valentijn, 
E.~A.\ 1989, \apss, 156, 127 
\bibitem[Peletier et al.(1990)]{Pe90a} Peletier, R.~F., Davies, R.~L., 
Illingworth, G.~D., Davis, L.~E., \& Cawson, M.\ 1990, \aj, 100, 1091
\bibitem[Peletier et al.(1995)]{Pe95} Peletier, R.~F., Valentijn, E.~A., 
Moorwood, A.~F.~M., Freudling, W., Knapen, J.~H., \& Beckman, J.~E.\ 1995, 
\aap, 300, L1
\bibitem[Peletier \& Balcells(1996)]{PB96} Peletier, R.~F., \& Balcells, M.\ 
1996, \aj, 111, 2238
\bibitem[Peletier et al.(1999)]{Pe99} Peletier, R.~F., Balcells, M., Davies, 
R.~L., Andredakis, Y., Vazdekis, A., Burkert, A., \& Prada, F.\ 1999, \mnras, 
310, 703
\bibitem[Peletier et al.(2007)]{Pe07} Peletier, R.~F., et al.\ 2007, \mnras, 
379, 445
\bibitem[Pierini(2002)]{Pi02} Pierini, D.\ 2002, \mnras, 330, 997
\bibitem[Searle et al.(1973)]{Se73} Searle, L., Sargent, W.~L.~W., \& Bagnuolo, 
W.~G.\ 1973, \apj, 179, 427
\bibitem[Silva \& Elston(1994)]{SE94} Silva, D.~R., \& Elston, R.\ 1994, \apj, 
428, 511
\bibitem[Skrutskie et al.(2006)]{Sk06} Skrutskie, M.~F., et al.\ 2006, \aj, 
131, 1163
\bibitem[Solanes et al.(2002)]{So02} Solanes, J.~M., Sanchis, T., 
Salvador-Sol{\'e}, E., Giovanelli, R., \& Haynes, M.~P.\ 2002, \aj, 124, 2440
\bibitem[Suh et al.(2010)]{Su10} Suh, H., Jeong, H., Oh, K., Yi, S.~K., 
Ferreras, I., \& Schawinski, K.\ 2010, \apjs, 187, 374
\bibitem[Tamura \& Ohta(2000)]{TO00} Tamura, N., \& Ohta, K.\ 2000, \aj, 120, 
533
\bibitem[Tamura et al.(2000)]{Ta00} Tamura, N., Kobayashi, C., Arimoto, N., 
Kodama, T., \& Ohta, K.\ 2000, \aj, 119, 2134
\bibitem[Tamura \& Ohta(2003)]{TO03} Tamura, N., \& Ohta, K.\ 2003, \aj, 126, 
596
\bibitem[Taylor et al.(2005)]{Ta05} Taylor, V.~A., Jansen, R.~A., Windhorst, 
R.~A., Odewahn, S.~C., \& Hibbard, J.~E.\ 2005, \apj, 630, 784
\bibitem[Terlevich et al.(1981)]{Te81} Terlevich, R., Davies, R.~L., Faber, 
S.~M., \& Burstein, D.\ 1981, \mnras, 196, 381
\bibitem[Terndrup et al.(1994)]{Te94} Terndrup, D.~M., Davies, R.~L., Frogel, 
J.~A., Depoy, D.~L., \& Wells, L.~A.\ 1994, \apj, 432, 518
\bibitem[Tinsley \& Gunn(1976)]{TG76} Tinsley, B.~M., \& Gunn, J.~E.\ 1976, 
\apj, 203, 52
\bibitem[Tortora et al.(2010)]{To10a} Tortora, C., Napolitano, N.~R., Cardone, 
V.~F., Capaccioli, M., Jetzer, P., \& Molinaro, R.\ 2010, \mnras, 407, 144
\bibitem[Trager et al.(2000)]{Tr00} Trager, S.~C., Faber, S.~M., Worthey, G., 
\& Gonz{\'a}lez, J.~J.\ 2000, \aj, 119, 1645
\bibitem[Tremonti et al.(2004)]{Tr04} Tremonti, C.~A., et al.\ 2004, \apj, 613, 
898
\bibitem[Tamm \& Tenjes(2006)]{TT06} Tamm, A., \& Tenjes, P.\ 2006, \aap, 449, 67
\bibitem[Tully et al.(1996)]{Tu96} Tully, R.~B., Verheijen, M.~A.~W., Pierce, 
M.~J., Huang, J.-S., \& Wainscoat, R.~J.\ 1996, \aj, 112, 2471
\bibitem[Vader et al.(1988)]{Va88} Vader, J.~P., Vigroux, L., Lachieze-Rey, M., 
\& Souviron, J.\ 1988, \aap, 203, 217
\bibitem[van Zee et al.(2004)]{vZ04} van Zee, L., Barton, E.~J., \& Skillman, E.~D.\ 2004, \aj, 128, 2797
\bibitem[West et al.(2009)]{We09b} West, A.~A., Garcia-Appadoo, D.~A., 
Dalcanton, J.~J., Disney, M.~J., Rockosi, C.~M., \& Ivezi{\'c}, {\v Z}.\ 2009, 
\aj, 138, 796
\bibitem[Worthey et al.(1992)]{Wo92} Worthey, G., Faber, S.~M., \& Gonzalez, 
J.~J.\ 1992, \apj, 398, 69
\bibitem[Worthey(1994)]{Wo94} Worthey, G.\ 1994, \apjs, 95, 107
\bibitem[Wu et al.(2005)]{Wu05} Wu, H., Shao, Z., Mo, H.~J., Xia, X., \& Deng, 
Z.\ 2005, \apj, 622, 244
\bibitem[Zaritsky et al.(1994)]{Za94} Zaritsky, D., Kennicutt, R.~C., Jr., \& 
Huchra, J.~P.\ 1994, \apj, 420, 87
\end{thebibliography}
\end{document}